\documentclass[twocolumn,aps,prb,superscriptaddress]{revtex4-1}
\usepackage{amssymb,amsmath}
\usepackage{graphics}

\usepackage{hyperref}
\usepackage{braket}

\providecommand{\abs}[1]{\vert#1\vert}

\providecommand{\e}{\text{e}}

\begin{document}
\title{Quantum thermodynamics of the driven resonant level model}

\author{Anton Bruch}
\author{Mark Thomas}
\author{Silvia Viola Kusminskiy}
\author{Felix von Oppen}

\affiliation{\mbox{Dahlem Center for Complex Quantum Systems and Fachbereich Physik, Freie Universit\"at Berlin, 14195 Berlin, Germany}}
\author{Abraham Nitzan}
\affiliation{\mbox{Dahlem Center for Complex Quantum Systems and Fachbereich Physik, Freie Universit\"at Berlin, 14195 Berlin, Germany}}
\affiliation{Department of Chemistry, University of Pennsylvania, Philadelphia, PA, 19104, USA}
\affiliation{School of Chemistry, Tel Aviv University, Tel Aviv 69978, Israel}

\date{\today}
\begin{abstract}
We present a consistent thermodynamic theory for the resonant level model in the wide band limit, whose level energy is driven slowly by an external force. The problem of defining 'system' and 'bath' in the strong coupling regime is circumvented by considering as the 'system' everything that is influenced by the externally driven level. The thermodynamic functions that are obtained to first order beyond the quasistatic limit fulfill the first and second law with a positive entropy production, successfully connect to the forces experienced by the external driving, and reproduce the correct weak coupling limit of stochastic thermodynamics.

\end{abstract}

\maketitle


\section{Introduction}

Classical machines such as heat engines and refrigerators are described by thermodynamic laws which characterize the processes by which a subsystem exchanges energy -- in the form of both heat and work -- and particles with its environment. With advances in nanofabrication, corresponding devices can now be realized on smaller and smaller scales (see Ref.\ \onlinecite{Kay2015} and references therein), demanding an extension of the thermodynamic description to the nanoscale. 

Such an extension poses several fundamental issues. Perhaps the most pressing among these is the proper accounting for the system-bath coupling. Macroscopic thermodynamics and statistical mechanics are based on models that describe systems whose intensive properties are governed by their coupling to equilibrium reservoirs. The vanishing surface-to-volume ratio in the thermodynamic limit  justifies the practice of disregarding the system-bath coupling in the thermodynamic description. In contrast, when considering the thermodynamics of small systems special attention has to be given to both the definition of the 'system' and consequently the 'bath', and their mutual interaction.\footnote{These issues are not limited to the quantum regime. However, the treatment of ultrasmall systems frequently requires quantum considerations.} 

These issues have been the subject of several recent papers, which address systems such as the resonant level model considered here \cite{LilianaDynamical,EspositoQThermo} or quantum particles strongly coupled to a harmonic oscillator bath.\cite{Allahverdyan2000a,Hanggi2008,Hilt2009a} Recent work has also addressed fluctuation theorems which characterize the stochastic behavior of thermodynamic quantities in quantum systems that are strongly coupled to their environment \cite{Campisi2009} and the efficiencies of different energy converting processes in quantum thermoelectric devices.\cite{Esposito2015,Ludovico2015} 

One outstanding issue is the need to derive a consistent formulation for the non-equilibrium thermodynamics of such strongly coupled system. This requires proper accounting of energy conservation as well a proper definition of entropy that will lead to entropy production consistent with the second law of thermodynamics. In particular, the entropy production is the central element in deriving efficiencies for various energy-conversion processes and characterizes the irreversibility of the process. It is thus an essential aspect of the non-equilibrium thermodynamics of nanoscale devices.\cite{Esposito2010a,Deffner2011a}

In this work, we study the thermodynamics of the driven resonant level model. This non-interacting model describes a single spinless electronic level (say, of a quantum dot) coupled to one or more leads described as free-electron metals. This system has long been studied as the simplest model for conducting nanoscopic junctions involving molecular or quantum dot bridges. When the resonant level energy and/or the level-lead coupling are driven by an external agent such as a gate voltage, it becomes a model for a quantum nano-engine, for which the above issues can be investigated. Our goal is to formulate a consistent non-equilibrium thermodynamic theory that will hold beyond the quasi-static limit in which the system remains in equilibrium and strictly follows the driving adiabatically.

Finding a consistent thermodynamic description of this model is non-trivial.\cite{LilianaDynamical,EspositoQThermo} First, the level-lead coupling itself has to be accounted for. Second, the strong hybridization of the dot level with the lead electronic states makes it necessary to develop an energy-resolved (or quantum) description of the dynamic processes, which goes beyond the kinetic (master-equation) schemes and stochastic approaches that are usually derived in the weak-coupling (or classical) limit.

Esposito et al.\cite{EspositoQThermo} pointed out these difficulties and, addressing the general case (i.e., including the driving in both the level energy and the level-lead coupling), formulated the basic laws of thermodynamics in a manner which includes the effects of irreversible driving through a modified spectral density. While satisfying the laws of thermodynamics, this formulation does not yield the known equilibrium forms of these thermodynamic functions in the quasistatic limit, already in the wide band limit and for time-independent level-lead coupling. 

Here we present an alternative formulation of the non-equilibrium thermodynamics of the driven resonant level model, albeit for the more restricted case where the driving affects only the level energy. In developing a consistent thermodynamic description of this model, we are guided by several basic requirements: The thermodynamic functions must (i) reduce to the correct quasistatic (equilibrium) limit, (ii) fulfill particle and energy conservation at each order, (iii) predict a positive entropy production reflecting the irreversibility of the transformations, and (iv) correctly connect to the forces experienced by the driving (see Refs.\ \onlinecite{BodePRB} and \onlinecite{bode2012current} for a general discussion and calculations of these forces). In departure from attempts to address the thermodynamic functions of the dot itself, which are marred by the need for a proper partitioning of the dot-lead coupling between the various subsystems, \cite{LilianaDynamical,EspositoQThermo,Esposito2014} we focus on 
the changes in the thermodynamic properties of the overall system (dot and lead) which result from local changes in parameters (i.e., the energy of the resonant level in the present context). This circumvents the need to address the contribution of the system-bath coupling to the thermodynamic functions of the dot, and instead defines the 'system' as that part of the 'world' which is influenced by the dynamics of the externally driven resonant level. We will henceforth refer to this part of the overall system as the extended resonant level.\footnote{ Note that, as we work in the grand canonical ensemble framework, the metal lead in our 'world' is assumed to be weakly open to an equilibrium bath of given temperature and electronic chemical potential.}

This paper is organized as follows. In Sec.\ \ref{Model}, we introduce the model. Section \ref{Equ.Thermodyn} contains a derivation of the equilibrium thermodynamics of the extended resonant level from the grand potential. Section \ref{NonEqu.Thermodyn} extends these thermodynamic functions to finite driving speed. To this end, we start with their representations in terms of quasistatic expectation values of operators, obtained in Sec.\ \ref{Equ.Thermodyn}, and expand these to linear order in the driving speed. This is done using a gradient expansion within the framework of non-equilibrium Green's functions. In Sec.\ \ref{ClassLimit}, we show that for weak level-lead coupling, our theory approaches the expected classical Master equation limit. We conclude in Sec.\ \ref{sec:Concl}. We have relegated most explicit calculations to a series of appendices in order not to break the flow of the main arguments. 

\section{Model}\label{Model}

We consider a single localized electronic level coupled to a free electron metal at temperature $T$ and chemical potential $\mu$. The Hamiltonian of the full system is
\begin{equation} \label{H}
 H=H_{D}+H_{V}+H_{B}\, ,
\end{equation}
where $H_{D}$, $H_{B}$, and $H_{V}$ denote the Hamiltonians of the dot, 
\begin{equation}
  H_{D}=\varepsilon_{d}(t)d^{\dagger}d\, ,\label{eq:Hd}
\end{equation}
of the metal lead,
\begin{equation}
H_{B}=\sum_{k}\varepsilon_{k}{c}_{k}^{\dagger}{c}_{k}\, ,\label{eq:HK}
\end{equation}
and of the lead-dot coupling,
\begin{equation}
H_{V}=\sum_{k}\left(V_{k}d^{\dagger}c_{k}+\text{h.c.}\right)\, .\label{eq:HV}
\end{equation}
Here, $d$ annihilates an electron in the dot level, $c_k$ an electron with momentum $k$ and energy $\varepsilon_k$ in the lead, and $V_k$ denotes the coupling strength between dot level and lead.  

The dot energy $\varepsilon _{d} \left(t\right)$ is driven by an external force. Our goal is to elucidate the effect of this driving on the thermodynamic properties of the system. We limit ourselves to the simplest situation of a single driven dot level, a single macroscopic lead, and the wide band approximation. (Alternative coupling models, see, e.g. Ref.\ \onlinecite{Ajisaka2012}, can be considered.) Apart from the driving, the lead is assumed to be in thermal equilibrium characterized by a temperature $T$ and an electronic chemical potential $\mu$. In the wide band approximation the retarded dot self-energy
\begin{equation} 
\Sigma ^{R} (\varepsilon)=\lim _{\eta \to 0} \sum _{k}\frac{\left|V_{k} \right|^{2} }{\varepsilon -\varepsilon _{k} +i\eta }  = -\frac{i}{2} \Gamma \,\label{SigmaR}
\end{equation} 
can be taken as purely imaginary and energy independent for energies $\varepsilon$ well within the bandwidth of the lead and vanishes for energies outside the band (see Appendix \ref{WBL}). It is furthermore proportional to the decay rate of the dot electrons into the lead $\Gamma = 2\pi \sum _{k}\left|V_{k} \right|^{2} \delta \left(\varepsilon -\varepsilon _{k} \right)$. 
Consequently, the spectral function associated with the dot's electronic state is a Lorentzian of width $\Gamma $ centered at $\varepsilon _{d}$,
\begin{equation} \label{ZEqnNum204971} 
A\left(\varepsilon \right)=\frac{\Gamma  }{\left(\varepsilon -\varepsilon _{d} \right)^{2} +\left({\Gamma  \mathord{\left/ {\vphantom {\Gamma/  2}} \right. \kern-\nulldelimiterspace} 2} \right)^{2} } \,.  
\end{equation} 
The broadening necessitates an energy resolved description of the electronic response to changes in the level energy and is responsible for the quantum nature of the problem. In Sec.\ \ref{ClassLimit} we show that our quantum results reduce to their classical counterparts in the limit $\Gamma \ll k_{B} T$ ($k_{B} $ is the Boltzmann constant). As already mentioned,  strong hybridization of dot and lead results in a reaction of the lead to  changes in the level energy. This makes the definition of thermodynamic quantities associated with the driven subsystem alone a difficult task. We overcome this problem by considering as the driven system the entire part of the 'world' that is affected by changes in the dot level, as shown in the next section. 

\section{Equilibrium Thermodynamics}\label{Equ.Thermodyn}

When $\varepsilon _{d}(t)$ moves infinitely slowly, the change induced by the driving is quasistatic and reversible.\footnote{The velocity of the level is measured by $\dot{\varepsilon}_d / \Gamma $ and the detailed condition for the process being quasistatic depends on whether $k_B T< \Gamma$ or $k_B T> \Gamma $. In these limits, one obtains the conditions $\dot{\varepsilon}_{d} / \Gamma \ll\Gamma $ and $\dot{\varepsilon}_{d} / \Gamma \ll k_B T $, respectively.} The system stays in equilibrium at all times and follows the change in $\varepsilon _{d}$ adiabatically. The desired thermodynamic functions can then be calculated from equilibrium thermodynamics. We do this in the grand canonical framework, where our 'full' system (i.e., dot and lead) is coupled to a reservoir that controls its temperature $T=k_{B} \beta ^{-1} $ and chemical potential $\mu $. 
In the free electron model, the grand partition function $\Xi $ and the grand potential $\Omega =-k_{B} T\ln \Xi $ can be evaluated exactly, yielding
\begin{equation} \label{OmegaTot} 
\Omega _{\rm tot} =-k_{B} T\int \frac{d\varepsilon }{2\pi } \rho \left(\varepsilon \right)\ln \left(1+e^{-\beta \left(\varepsilon -\mu \right)} \right)\, , 
\end{equation} 
where the label `tot' stands for this being the grand potential of the total system. We emphasize that the total system comprises everything that is described by the Hamiltonians (\ref{eq:Hd})-(\ref{eq:HV}), namely the dot, the lead, and their coupling. In Eq.\ \eqref{OmegaTot}, $\rho \left(\varepsilon \right)$  is the density of states of the system as given by the trace of the spectral function,
\begin{equation}  
 \rho \left(\varepsilon \right)=\sum _{n} \, A_{nn} (\varepsilon ) \,.
\end{equation} 
Here,  $A_{nn} (\varepsilon )=-2\, \text{Im}\, G_{nn}^{R} (\varepsilon )$ with the retarded Green's function 
\begin{equation}
 G_{nn'}^{R} (t,t')=-i\Theta (t-t')\left\langle \left\{c_{n} (t),c_{n'}^{\dag } (t')\right\}\right\rangle\, . 
\end{equation} 
The index $n$ enumerates all single-particle states (lead and dot). For better comparison with the recent work of Ref.\ \onlinecite{EspositoQThermo}, we present the calculation of the density of states beyond the wide band limit, which is shown in App.\ \ref{DensityOfStates}. The result is 
\begin{align} \label{RhoFull}
 \rho \left(\varepsilon \right)=&A_{dd} (\varepsilon )\left(1-\frac{d}{d\varepsilon } \, \text{Re}\, \Sigma ^{R} (\varepsilon )\right)\nonumber \\
 &+2\, \text{Re}\, G_{dd}^{R}(\varepsilon) \frac{d}{d\varepsilon } \,\text{Im}\, \Sigma ^{R} (\varepsilon )+\nu (\varepsilon )\, , 
\end{align} 
where $A_{dd}(\varepsilon )$ is the full spectral function associated with the dot's electronic state (i.e. not in the wide band limit), $\Sigma ^{R} $ is the corresponding retarded self-energy, and  $\nu \left(\varepsilon \right)$ is the density of states of the free lead. The $\varepsilon _{d}$-dependent term of the grand potential stems from the density of states $\rho \left(\varepsilon \right)$  and arises from the first three of the four terms in Eq.\ \eqref{RhoFull}. In the  wide band limit, the second and third terms on the right hand side of Eq.\ \eqref{RhoFull} vanish, and the $\varepsilon _{d}$-dependent part of the density of states $\rho_{\varepsilon_d}$ is given by the spectral function $A(\varepsilon)$, Eq.\ \eqref{ZEqnNum204971}. In the general (non-wide-band) case, the $\varepsilon_d$-dependent part of the density of states is similar to the modified spectral function proposed in Ref.\ \onlinecite{EspositoQThermo}, with the difference that the energy derivative in the third term is taken of 
the imaginary part of the self energy, while Esposito et al.\ have a contribution $-2\, \text{Im}\, \Sigma ^{R} \partial_\varepsilon \, \text{Re}\, G_{dd}^{R}(\varepsilon)$ to their modified spectral function. This leads to different thermodynamic functions calculated with the help of the density of states, also in the wide band limit. We refer to the $\varepsilon_d$-dependent part of the system as the extended resonant level, since it accounts for the change of the surrounding in response to changing the level energy. 

We now use the $\varepsilon _{d}$-dependent part of the density of states  $\rho_{\varepsilon_d} \left(\varepsilon \right)=A(\varepsilon)$ to calculate the $\varepsilon _{d}$-dependent contribution to the grand potential $\Omega $, which in turn yields the corresponding $\varepsilon _{d}$-dependent contributions to all the thermodynamic functions of the system. In particular, we calculate the entropy $S^{(0)} $, the internal energy $E^{(0)}$, and the particle number $N^{(0)}$ of the extended resonant level in equilibrium, i.e., for a frozen dot level, and show how they evolve when the dot level is changed quasistatically by an external force. We use superscripts on the thermodynamic functions to indicate to which order in the level velocity $\dot{\varepsilon}_d$ they are calculated. Furthermore we show how these quantities can be represented, in the model considered, as quasistatic expectation values of operators. This observation provides a convenient route for extending the quasistatic thermodynamic 
quantities to 
non-equilibrium, i.e., to situations where the dot level is moved at finite speed (see Sec.\ \ref{NonEqu.Thermodyn}). 

In the following, the notation $\Omega$, $S^{(0)}$,  $E^{(0)}$, $N^{(0)}$ and the corresponding names grand potential, entropy, energy, and particle number always refer to the $\varepsilon _{d}$-dependent parts of these functions. The grand potential takes the form 
\begin{equation} \label{GrindEQ__9_} 
\Omega =-k_{B} T\int \frac{d\varepsilon }{2\pi } A\ln \left(1+e^{-\beta \left(\varepsilon -\mu \right)} \right)\, . 
\end{equation} 
Here and in the following, we omit energy arguments for better readability. The particle number, entropy and energy are given by 
\begin{align} 
N^{(0)} &=-\frac{\partial \Omega }{\partial \mu } =\int \frac{d\varepsilon }{2\pi } A\,f\, ,\label{GrindEQ__10_} \\ 
S^{(0)} & = -\frac{\partial\Omega}{\partial T} \nonumber \\
 &= k_B\,\int\frac{d\varepsilon}{2\pi}A\left[ \beta\left(\varepsilon-\mu\right)f+\ln\left(1+\e^{-\beta(\varepsilon-\mu)}\right) \right] \nonumber \\
&=  k_{B}\int\frac{d\varepsilon}{2\pi}A\left[-f\ln f-\left(1-f\right)\ln\left(1-f\right)\right]\,, \label{S_0}
\end{align}
and
\begin{equation} \label{GrindEQ__11_} 
  E^{(0)} =\Omega +\mu N^{(0)}  + T S^{(0)} =\int \frac{d\varepsilon }{2\pi } \varepsilon A\,f\,,  
\end{equation} 
where $f$ is the Fermi-Dirac distribution. In the wide band limit, the grand potential as well as the internal energy depend on the bandwidth $D$ and diverge in the limit $D\to\infty$. However, this only affects the reference point from which the grand potential and the internal energy are measured. Here, we are interested in the thermodynamic relations between \textit{changes} in these quantities as the dot level $\epsilon_d$ varies. These changes converge to bandwidth-independent values in the limit of an infinite bandwidth (see the detailed discussion in App.\ \ref{WBL}).

Equation \eqref{GrindEQ__10_} implies that, in the wide band limit, the $\varepsilon _{d}$-dependent part of the equilibrium particle number $N^{(0)} $ is given by the quasistatic dot occupation $N^{(0)} =\left\langle d^{\dag } (t)d(t)\right\rangle ^{(0)}$, namely the equilibrium occupation for the instantaneous value of $\varepsilon _{d}$ . The contribution to the energy, Eq.\ \eqref{GrindEQ__11_}, explicitly shows that the coupling to the environment affects the energy cost associated with changes of the bare dot energy $\varepsilon _{d}$, as it cannot be represented as an expectation value of $H_D$ only. Equation \eqref{S_0} is the energy resolved version of the Gibbs entropy of a single fermionic level with equilibrium occupation probability $f$, weighted by the spectral function of the dot electrons. For $T\to 0$, the term in square brackets in Eq.\ \eqref{S_0} for $S^{(0)}$ tends to zero for $\varepsilon \neq \mu$  and to $\ln2 $ for $\varepsilon = \mu$, 
reflecting the degeneracy at the Fermi edge. Integrating over energy leads to a vanishing equilibrium entropy $S^{(0)}$  of the extended resonant level for  $T\to 0$. 

It is important to note that the equilibrium energy of the extended resonant level, namely the $\varepsilon_d$-dependent part of the total (dot plus lead) internal energy, can be expressed as a sum of contributions from the different terms in the Hamiltonian \eqref{H}. In particular, as shown in App.\ \ref{IntEnergyAp}, the part of the internal energy $E^{(0)} $ given by Eq.\ \eqref{GrindEQ__11_} can be represented by the quasistatic expectation value $E^{(0)}=\braket{H_{D}}^{(0)}+\frac{1}{2}\braket{H_{V}}^{(0)} $. This appears to indicate that, in the model considered, half the energy associated with the coupling $ H_{V} $ can be  attributed to the extended resonant level. This interpretation, however, is an oversimplification as may be realized from the following: Calculating the $\varepsilon_d$-dependent part of the averages of $H_D$, $H_V$, and $H_B$ from the grand potential, Eq.\ \eqref{GrindEQ__9_}, we obtain $\braket{H_B}_{\varepsilon_d}=-\int\frac{d\varepsilon}{2\pi}\left(\varepsilon-\varepsilon_{d}
\right)\, A\,f$, $\braket{H_V}_{\varepsilon_d}=2\int\frac{d\varepsilon}{2\pi}(\varepsilon-\varepsilon_{d})A\,f$, and  $\braket{H_D}_{\varepsilon_d}=\varepsilon_d\int\frac{d\varepsilon}{2\pi}A\,f$ (see App.\ \ref{IntEnergyAp}). It is interesting to note that not only $\braket{H_V}$ but also $\braket{H_B}$ has an $\varepsilon_d$-dependent part and together with $\braket{H_D}$ they add up to $E^{(0)}$, Eq.\ \eqref{GrindEQ__11_}. In fact, the contributions of $H_V$ and $H_B$ add to $\braket{H_B}_{\varepsilon_d}+\braket{H_V}_{\varepsilon_d}= \frac{1}{2}\braket{H_{V}}^{(0)}$, which shows the intricate physical origin of the symmetric splitting.

An apparent symmetric splitting of the coupling energy in the wide band limit of the resonant level model between an effective driven system $H_{D} +{\tfrac{1}{2}} H_{V} $ and an effective bath $H_{B} +{\tfrac{1}{2}} H_{V} $  was also found in the case of periodic driving.\cite{LilianaDynamical} It should be emphasized that this separation, namely assigning parts of the calculated thermodynamic functions to the different subsystems is not needed in the present analysis of the equilibrium thermodynamics. We allude to it both because it has been considered in recent discussions \cite{LilianaDynamical} and because it can help building intuition about the system behavior. Furthermore it serves as a convenient starting point for the Green's function based calculation of the internal energy when the level moves at finite velocity.

Next, we consider the evolution of the thermodynamic functions when changing the dot level quasistatically. In particular, we examine the different contributions to the reversible energy change $dE^{(0)}$, the reversible work $dW^{(0)}$, the heat $dQ^{(0)}$, and the chemical work $\mu \, dN^{(0)}$. These satisfy energy conservation as expressed by the first law,
\begin{equation} \label{FirstLaw} 
 dE^{(0)} =dW^{(0)} +dQ^{(0)} +\mu \, dN^{(0)} \,,
\end{equation}
when applied to the extended resonant level. Note that this equation relates properties of the full system (dot + lead). But because the individual terms result from changes in the bare dot energy $\varepsilon _{d} $, they are often referred to as changes in the corresponding dot property.

The reversible work is given by the change in the grand potential upon changing the level energy, $dW^{(0)} =d\varepsilon _{d} \, \partial_{\varepsilon_d} \Omega$. Expressed as an equation for the power $\dot{W}^{(1)}$, this takes the form 
\begin{equation} \label{GrindEQ__13_} 
 \dot{W}^{(1)} =\dot{\varepsilon }_{d} N^{(0)} \left(\varepsilon _{d} \right)\, =\dot{\varepsilon }_{d} \left\langle d^{\dag } (t)d(t)\right\rangle ^{(0)}  \,.
\end{equation} 
It is frequently the case that the time dependence of $\varepsilon _{d} \left(t\right)$ reflects the dynamics of some external coordinate, $\varepsilon _{d} \left(t\right)=Mx_{d} \left(t\right)$ with a coupling parameter \textit{M}. The quantity  $F=-M\left\langle d^{\dag } (t)d(t)\right\rangle^{(0)} $ is then the quasistatic force needed to change the level energy. General expressions for such forces were obtained in the context of adiabatic reaction forces.\cite{BodePRB,bode2012current} 

The quasistatic heat leaving or entering the system is calculated from $dQ^{(0)} =Td\varepsilon _{d} \, \partial_{\varepsilon_d} S^{(0)} $, with $S^{(0)}$ given by Eq.\ \eqref{S_0}. By noting that $A(\varepsilon) $ depends only on $(\varepsilon-\varepsilon_d) $ and integrating by parts, the corresponding quasistatic heat current takes the form 
\begin{equation} \label{GrindEQ__14_} 
 \dot{Q}^{(1)} =T\dot{\varepsilon }_{d} {\frac{\partial S^{(0)} }{\partial \varepsilon _{d} }} =\dot{\varepsilon }_{d} \int \frac{d\varepsilon }{2\pi }  \left(\varepsilon -\mu \right)A\partial_\varepsilon f.
\end{equation} 
With $N^{(0)}$ in Eq.\ \eqref{GrindEQ__10_}, the quasistatic particle current $\dot{N}^{(1)}=\dot{\varepsilon }_{d} \partial_{\varepsilon _{d}} N^{(0)}$ is  given by 
\begin{equation} \label{GrindEQ__15_} 
 \dot{N}^{(1)} =\dot{\varepsilon }_{d} \int \frac{d\varepsilon }{2\pi } A\partial_\varepsilon f \, . 
\end{equation} 
The quasistatic change in the system's energy associated with the change in $\varepsilon_{d} $ is given by
\begin{equation} \label{GrindEQ__16_} 
 \dot{E}^{(1)} =\dot{\varepsilon }_{d} \frac{\partial E^{(0)} }{\partial \varepsilon _{d} } =\dot{\varepsilon }_{d} \int \frac{d\varepsilon }{2\pi } \varepsilon \frac{\partial A }{\partial \varepsilon _{d} } f
\end{equation} 
and is easily seen to indeed satisfy the first law, Eq.\ \eqref{FirstLaw}, since $\dot{E}^{(1)} =\dot{W}^{(1)} +\dot{Q}^{(1)} +\mu \, \dot{N}^{(1)} $. Note that the quasistatic power $\dot{W}^{(1)}$, the currents $\dot{N}^{(1)}$ and $\dot{Q}^{(1)} $, and the rate of energy change $\dot{E}^{(1)} $ are \textit{linear} in the driving speed, as indicated by the superscript.

We end our discussion of quasistatic (equilibrium) processes with several comments:

(a) The integrand of $\dot{N}^{(1)} $ can be understood as an energy resolved particle current $J^{\left(1\right)} (\varepsilon )=\dot{\varepsilon }_{d} A \partial_\varepsilon f $ and the right hand side of Eq.\ \eqref{GrindEQ__14_} can be expressed in terms of the same current 
\begin{equation} \label{GrindEQ__17_} 
 \dot{Q}^{(1)} =\int \frac{d\varepsilon }{2\pi }J^{\left(1\right)} (\varepsilon )\left(\varepsilon -\mu \right)\, . 
\end{equation} 
Consequently, $J_{Q}^{\left(1\right)} \left(\varepsilon \right)= J^{\left(1\right)} \left(\varepsilon \right)\left(\varepsilon -\mu \right)$ can be identified as the energy resolved heat current, providing physical insight into the nature of this current. It is important to note that identifying the integrand of an energy integral such as the particle current $\dot{N}^{(1)}$ in Eq.\ \eqref{GrindEQ__15_} as an energy resolved current is open to ambiguity. Other expressions could also be chosen following integration by parts. Considering the particle and heat currents together serves to resolve this ambiguity. 

(b) For quasistatic processes, we could calculate the particle, energy, and heat currents without assigning these variables to expectation values of the dot operators themselves. Especially the quasistatic heat current, Eq.\ \eqref{GrindEQ__14_}, was obtained without relying on any specific forms for the energetic properties of the dot itself. In particular the symmetric splitting of the coupling Hamiltonian between dot and lead, discussed above, was not used. It can, however, also be calculated from expectation values using the symmetric splitting into effective bath and system introduced above. Indeed, we show in App.\ \ref{HeatFluxAp} that to lowest order in the level speed, the adiabatic heat current $\dot{Q}^{(1)}$ given in Eq.\ \eqref{GrindEQ__14_} is reproduced by the change of the energy of the effective bath $H_{B} +{\tfrac{1}{2}} H_{V} $ minus the chemical contribution of the particle flow, 
\begin{equation} \label{HeatSymm} 
\dot{Q}^{(1)}=-\frac{d}{dt} \left\langle H_{B} +\frac{1}{2} H_{V} \right\rangle ^{(0)} -\mu \frac{d}{dt} N^{(0)} \, . 
\end{equation} 
Eq.\ \eqref{HeatSymm} confirms, for the present model and the wide band limit, the consistency of the symmetric splitting of the coupling Hamiltonian $H_V$ into an effective bath and an effective driven system. This will serve as a convenient starting point for the calculation of the heat current at finite level speed. Note, however, that for more general models (e.g., beyond the wide band approximation and with variations in the level-lead coupling), the possibility to express the change in thermodynamic variables in terms of expectation values of ``system operators" is an open problem and subject to several difficulties.\cite{Esposito2014}

(c) In the quasistatic process, the entropy change $\dot{S}^{(1)} =\dot{\varepsilon }_{d} \, \partial_{\varepsilon_d}S^{(0)}$ is given by the corresponding heat current, $\dot{Q}^{(1)} =T\dot{S}^{(1)} $, indicating that no entropy is produced. This is not the case when the level moves at finite speed and dissipation sets in, as discussed in the next section.

We have described the equilibrium thermodynamics of the resonant level model and calculated the reversible change of the thermodynamic quantities in the wide band limit. We represented all thermodynamic quantities of the extended resonant level as quasistatic expectation values of operators. Next we extend our discussion to the non-adiabatic regime and consider the effect of moving the dot level energy at a small, but finite speed. 

\section{Non-equilibrium Thermodynamics}\label{NonEqu.Thermodyn}

In this section, we consider the changes in thermodynamic quantities when the dot level moves at finite speed. For this non-equilibrium process we cannot use the equilibrium grand potential as a starting point. Instead, we extend our quasistatic results to finite speed processes by expanding the expectation values of the operators associated with the thermodynamic variables in powers of the level velocity, using the non-equilibrium Green's function approach together with the gradient expansion in the Wigner representation. Our theory should follow three guidelines: First, all non-equilibrium quantities should converge to their equilibrium forms, obtained in the previous section, in the limit of vanishing speed. Second, higher order corrections should satisfy conservation of energy and particle number at the corresponding order. Third, the non-equilibrium entropy of the extended resonant level should lead to positive entropy production characterizing the irreversibility of the process. Note that the 
corrections obtained below are of different orders in the level speed. The corrections to the equilibrium values of the thermodynamic variables themselves are linear in $\dot{\varepsilon}_d$, while the correction to their fluxes are quadratic. The corresponding order is again indicated by the superscript assigned to the different variables. We also assume a linear motion of the dot level, $\ddot{\varepsilon}_d=0$.

\textit{Particle number}. We extend the calculation of the particle number of the resonant level to finite speed by expanding the lesser Green's function $\left\langle d^{\dag } (t)d(t)\right\rangle =-iG_{dd}^{<} (t,t)$ to linear order in the level speed. This is done in App.\ \ref{NEGF}. Alternatively, the effect of the level speed on the dot occupation can be expressed through a non-equilibrium distribution function $\phi$ (as done in Ref.\ \onlinecite{EspositoQThermo}), which is related to the Wigner transform of the lesser Green's function via $G^{<} =iA\phi $. The equation of motion for $\phi$ and its solution are given in App.\ \ref{QBoltzmann}, and the final result for the non-equilibrium distribution $\phi$ is
\begin{equation}\label{phi}
 \phi=f-\frac{\dot{\varepsilon}_d}{2} \partial_\varepsilon f A\, .
\end{equation}
Both approaches are equivalent and lead to $G^{<} =iA \left(f-\frac{\dot{\varepsilon}_{d}}{2}\partial_{\varepsilon}f\, A \right) $ and therefore to a correction to the particle number linear in the velocity, 
\begin{equation} \label{GrindEQ__19_} 
N^{(1)} =-\frac{\mathop{\dot{\varepsilon }}_{d} }{2} \int \frac{d\varepsilon }{2\pi } \partial_\varepsilon f \, A^{2} \, . 
\end{equation} 
This correction in the particle number accounts for the fact that the dot population lags behind the equilibrium value since electrons are not exchanged fast enough with the leads. The time derivative of Eq.\ \eqref{GrindEQ__19_} now yields the correction $\dot{N}^{(2)} =\frac{d}{dt} N^{(1)}$ to the quasistatic current, $\dot{N}^{(1)}$, that takes the form
\begin{equation} \label{GrindEQ__20_} 
\dot{N}^{(2)} =-\frac{\dot{\varepsilon }_{d}^2 }{2} \int \frac{d\varepsilon }{2\pi } \partial _{\varepsilon }^{2} f\, A^{2} \, . 
\end{equation} 
One might be tempted to identify the integrand of $\dot{N}^{(2)}$ as the second order correction to the energy resolved particle current. However, this cannot be done unambiguously because other expressions can be obtained after integration by parts. As before, more information can be obtained by considering the particle and heat currents together as further discussed below.  

\textit{Work}. The quasistatic work per unit time $\dot{W}^{(1)} =\dot{\varepsilon }_{d} N ^{(0)}$ can be extended to finite level speed with the correction to the dot occupation $N^{(1)} $, Eq.\ \eqref{GrindEQ__19_}. With this we readily obtain the extra power that the external driving has to provide for moving the level at finite speed by multiplying $N^{(1)} $, Eq.\ \eqref{GrindEQ__19_}, by the level speed
\begin{equation} \label{GrindEQ__22_} 
\dot{W}^{(2)} =-\frac{\mathop{\dot{\varepsilon }}_{d}^{2} }{2} \int \frac{d\varepsilon }{2\pi } \partial_\varepsilon f \, A^{2} \, . 
\end{equation} 
$\dot{W}^{(2)}$ thus corresponds to the power dissipated by driving the system at finite speed. When considering the time dependence of $\varepsilon _{d} \left(t\right)$ as reflecting the  dynamics of some external coordinate, $\varepsilon _{d} \left(t\right)=Mx_{d} \left(t\right)$, the dissipated power is caused by a friction force acting on the external coordinate $F_{\text{fric}}=-MN^{(1)}=-\gamma \dot{x}_{d}$. This yields the friction coefficient
\begin{equation} \label{friction} 
   \gamma =- \frac{M^2 }{2} \int \frac{d\varepsilon }{2\pi } \partial_\varepsilon f \, A^{2} \,.
\end{equation} 
The same expression for the friction in the resonant level model was found in Ref.\ \onlinecite{bode2012current}. 

\textit{Internal energy}. We showed above that the equilibrium internal energy of the extended resonant level can be represented as the quasistatic expectation value $E^{(0)}=\braket{H_{D}}^{(0)}+\frac{1}{2}\braket{H_{V}}^{(0)}$. Expanding the expectation values to first order in the velocity (see App.\ \ref{IntEnergyAp}), we obtain the first order correction to the internal energy,
 \begin{align}\label{IntEnergy1st}
E^{(1)}&=\frac{-\dot{\varepsilon}_{d}}{2} \int\frac{d\varepsilon}{2\pi}\varepsilon\partial_{\varepsilon}f A^2 \,.
\end{align}

\textit{Heat flux}. Taking the next order correction to the expression of the quasistatic heat flux, Eq.\ \eqref{HeatSymm}, in terms of the energy change in the effective bath and the chemical contribution (shown in App.\ \ref{HeatFluxAp})  gives the correction to the heat flux that originates from moving the level at finite speed,
\begin{equation} \label{GrindEQ__26_} 
\dot{Q}^{(2)}=-\frac{\dot{\varepsilon}_{d}^{2}}{2}\int\frac{d\varepsilon}{2\pi}(\varepsilon-\mu)\,\partial^{2}_\varepsilon f\, A^{2}\,.
\end{equation} 
As in the case of the quasistatic heat current, the integrand of the correction $\dot{Q}^{(2)} $ can be understood as heat $(\varepsilon -\mu )$ carried into the lead by the energy resolved particle current $J^{(2)} (\varepsilon )$, $\dot{Q}^{(2)}   =\int {\tfrac{d\varepsilon }{2\pi }} (\varepsilon -\mu )J^{(2)} (\varepsilon )$. The energy resolved particle current $J^{(2)} (\varepsilon )$ in turn is the properly chosen integrand in  ${\dot{N}}^{(2)} =\int {\tfrac{d\varepsilon }{2\pi }} J^{(2)} (\varepsilon )$ as given by Eq.\ \eqref{GrindEQ__20_}. This unambiguously defines the second order correction to the energy resolved particle current as $J^{(2)}= -\frac{\dot{\varepsilon}_{d}^{2}}{2}\partial^{2}_\varepsilon f\, A^{2}$.

\textit{Consistency checks.} The consistency of our thermodynamic description should be examined by its behavior in the quasistatic limit, by satisfying particle conservation, and by its adherence to the first law (energy conservation). Furthermore the entropy, discussed below, should give a consistent second law. Indeed, our expressions go over to the equilibrium (quasistatic) limit by construction, and taking the time derivative of the first order correction to the internal energy $E^{(1)}$, Eq.\ \eqref{IntEnergy1st}, shows (see App.\ \ref{IntEnergyAp})  that also the expressions for the first order corrections of particle number, internal energy, work, and heat satisfy the first law  $\dot{E}^{(2)} =\dot{W}^{\left(2\right)} +\dot{Q}^{\left(2\right)} +\mu \, \dot{N}^{\left(2\right)} $. 

As an additional check, we show in the following that the corrections to work, heat, and particle number exhibit the correct behavior under transformations between equilibrium points, corresponding to a path-independent change of internal energy and particle number. To this end, we consider a path between two points that essentially represent a system in equilibrium, namely the dot level $\varepsilon _{d}$ moving from a position far below $\mu$, where it is completely occupied, at time $t_{1}$ to a position far above $\mu$, where it is completely empty, at time $t_{2}$. The change of the particle number associated with this transformation is thus path-independent, requiring that the non-equilibrium correction $\dot{N}^{(2)}$  in Eq.\ \eqref{GrindEQ__20_} vanishes when integrated along this path 
\begin{equation} \label{GrindEQ__21_} 
  \Delta N^{(2)} =\int _{t_{1} }^{t_{2} } dt\dot{N}^{(2)} = 0 \,.
\end{equation} 
We show in App.\ \ref{particle conservation} that this is indeed the case. Furthermore, also the change in internal energy $\Delta E$ cannot depend on the path and must therefore be given by its adiabatic value, i.e., as an integral over time of $\dot{E}^{(1)}$ in Eq.\ \eqref{GrindEQ__16_}. This must hold although the instantaneous value of $E=E^{(0)}+E^{(1)}$ is velocity dependent, cp., Eq.\ \eqref{IntEnergy1st}. Thus, the extra work exerted for moving the level along this path at finite speed needs to appear as additional heat given to the leads, 
\begin{equation}
\int_{t_{1}}^{t_{2}}dt\dot{W}^{(2)}=-\int_{t_{1}}^{t_{2}}dt\dot{Q}^{(2)}\,.
\end{equation} 
We show in App.\ \ref{HeatWorkCompens} that this equality is indeed satisfied by the second order quantities Eqs.\ \eqref{GrindEQ__22_} and \eqref{GrindEQ__26_}. 

\textit{Entropy}. In addition to the consistency checks discussed above, the non-equilibrium correction to the entropy should comply with the second law of thermodynamics. A consideration of this issue requires a proper definition of the non-equilibrium entropy. In Sec.\ \ref{Equ.Thermodyn} we showed that the equilibrium entropy $S_0$ of the extended resonant level (cp., Eq.\ \eqref{S_0}) is an integral over the energy resolved version of the Gibbs entropy of a single fermionic level with equilibrium occupation probability $f$. In order to extend this result to finite level speeds, we follow Esposito et al.,\cite{EspositoQThermo} and use Eq.\ \eqref{S_0} as an ansatz for the non-equilibrium entropy after replacing the equilibrium distribution $f$ by its non-equilibrium counterpart $\phi$ given in Eq.\ \eqref{phi},
\begin{equation} \label{GrindEQ__28_} 
S=k_{B} \int \frac{d\varepsilon }{2\pi } A\left(-\phi \ln \phi -\left[1-\phi \right]\ln \left[1-\phi \right]\right)\, . 
\end{equation} 
Note that in contrast to Esposito et al.,\cite{EspositoQThermo} we define the non-equilibrium entropy with the standard broadened spectral function $A\left(\varepsilon\right)$ of the dot electrons. Consequently, our form of the non-equilibrium entropy smoothly connects to the equilibrium limit $S^{(0)}$ given in Eq.\ \eqref{S_0} above. Expanding Eq.\ \eqref{GrindEQ__28_} up to first order in $\dot{\varepsilon}_d$ leads to the form $S=S^{(0)}+S^{(1)}$, where $S^{(0)}$ is the equilibrium entropy Eq.\ \eqref{S_0} and  $S^{(1)}$ is the first order correction, 
\begin{equation} \label{S_1} 
S^{(1)}=\frac{-k_{B}\dot{\varepsilon}_d}{2}\int\frac{d\varepsilon}{2\pi}\left(\frac{\varepsilon-\mu}{k_{B}T}\right)\partial_\varepsilon f A^2\, . 
\end{equation} 
From Eq.\ \eqref{S_1} the correction to the entropy evolution (quadratic in the velocity) is given by
\begin{equation}
\dot{S}^{(2)}=\frac{\dot{\varepsilon}_{d}^{2}}{2T}\int\frac{d\varepsilon}{2\pi}\left(\varepsilon-\mu\right)\partial_{\varepsilon}f\,\partial_{\varepsilon}A^2\,.
\end{equation}
 While the change of the equilibrium entropy $\dot{S}_{0} =\dot{\varepsilon }_{d} \, \partial_{\varepsilon_d} S_{0} $ is solely given by the corresponding heat current, $\dot{Q}_{0} =T\dot{S}_{0} $, the second order correction $\frac{dS}{dt}^{(2)}$ cannot be written only in terms of the heat current $\dot{Q}^{(2)}/T$ in Eq.\ \eqref{GrindEQ__26_}. We identify the remaining entropy change as the entropy production $\dot{\mathbb{\mathcal{S}}}^{(2)}$,
\begin{equation}
\frac{dS}{dt}^{(2)}=\frac{\dot{Q}}{T}^{(2)}+\dot{\mathbb{\mathcal{S}}}^{(2)}\, .
\end{equation}
The entropy production can be related to the dissipated power, Eq.\ \eqref{GrindEQ__22_},  
\begin{equation}
\dot{\mathbb{\mathcal{S}}}^{(2)}=\frac{\dot{W}^{(2)}}{T}\geq0\,.
\end{equation}
Therefore the non-equilibrium entropy defined above obeys the second law of thermodynamics and the entropy production vanishes for quasistatic driving. Furthermore, the entropy production calculated for finite driving speeds is properly related to the dissipated power. We have thus found, for this model, a consistent extension of quantum thermodynamics to this non-equilibrium situation.

\section{Classical limit}\label{ClassLimit}
Here we show that the energy resolved thermodynamic quantities obtained above reduce to their classical equivalents in the limit $\Gamma\ll k_{B}T$. Thus, the quantum thermodynamics framework developed here is consistent with the familiar classical limit in which the dot level is well described by a Pauli master equation. The latter takes the form of a rate equation for the occupation probability of the resonant level $p$,
\begin{equation}\label{MasterEq}
\frac{dp}{dt}=-\Gamma\left[1-f(\varepsilon_{d})\right]p+\Gamma f(\varepsilon_{d})\left[1-p\right]\,.
\end{equation}
We first consider the thermodynamic implications of this dynamics. To this end, we solve Eq.\ \eqref{MasterEq}  to linear order in the velocity in terms of a static solution $f(\varepsilon_d) $ plus a velocity dependent correction,
\begin{equation}\label{ClassicalN}
p=N^{(0)}+N^{(1)}=f(\varepsilon_{d})-\frac{\dot{\varepsilon}_{d}f'(\varepsilon_{d})}{\Gamma}\,,  
\end{equation}
with $f'(\varepsilon_{d})=\partial_\varepsilon f \vert _ {\varepsilon_d}$. As in the strongly coupled quantum system considered above, the power that the external driving needs to provide is set by the dot occupation $\dot{W}=\dot{\varepsilon}_d\,N$. Eq.\ \eqref{ClassicalN} then directly yields the power up to second order as $\dot{W}^{(1)}+\dot{W}^{(2)}=\dot{\varepsilon}_d\,p$. In this weak coupling case, the $\varepsilon_{d}$-dependent part of the
thermodynamic properties of the overall system are well represented by those that are usually assigned to the dot itself. This leads directly to the classical internal energy, $E=\varepsilon_d\,N$, up to first order in the velocity
\begin{equation}
E^{(0)}+E^{(1)}=\varepsilon_d \left(f(\varepsilon_{d})-\frac{\dot{\varepsilon}_{d}f'(\varepsilon_{d})}{\Gamma}\right)\,,
\end{equation}
and to the heat flux between the dot and its environment, $\dot{Q}=\left(\varepsilon_{d}-\mu\right)\dot{N}$, up to second order in the velocity
\begin{equation}\label{QClassical}
\dot{Q}^{(1)}+\dot{Q}^{(2)}=(\varepsilon_{d}-\mu)\left(\dot{\varepsilon}_{d}f'\left(\varepsilon_{d}\right)-\frac{\dot{\varepsilon}^2_d}{\Gamma}f''(\varepsilon_d)\right)\,.
\end{equation}
Finally, the $\varepsilon_d$-dependent part of the entropy in this weak coupling limit is again given by the dot entropy itself. Assuming the latter is given by the Gibbs form 
\begin{equation}
S=-k_{B}\left(p\ln p+\left(1-p\right)\ln\left(1-p\right)\right)\,,
\end{equation}
one obtains
\begin{equation}\label{SdotClass}
\dot{S}^{(1)} =\frac{\dot{Q}^{(1)}}{T}\quad \text{and} \quad \dot{S}^{(2)} =\frac{\dot{Q}^{(2)}}{T}+\frac{\dot{W}^{(2)}}{T}\,, 
\end{equation}
where $\dot{W}^{(2)}=-\frac{\dot{\varepsilon}_{d}^{2}}{\Gamma}f'(\varepsilon_{d})$.

This weak coupling thermodynamics can be directly reproduced from the thermodynamic quantities of the resonant level model derived in Secs.\ \ref{Equ.Thermodyn} and \ref{NonEqu.Thermodyn} by taking the limit $\Gamma\ll k_{B}T$. In this limit, the spectral function $A$ becomes strongly peaked around $\varepsilon_{d}$ so that we can neglect the variation of the Fermi distribution within the broadened level and, in case the thermodynamic function contains the spectral function $A$ to the first power, replace it by a $\delta$-function, $A\rightarrow \delta(\varepsilon-\varepsilon_d)$. Expressions that contain higher powers of $A$ have to be handled more carefully by performing the integral over the spectral functions explicitly. Thus, for example, Eq.\ \eqref{GrindEQ__26_} leads to
\begin{align}
 \dot{Q}^{(2)}&=-\int\frac{d\varepsilon}{2\pi}(\varepsilon-\mu)\frac{\dot{\varepsilon}_{d}^{2}}{2}\partial_{\varepsilon}^{2}f\,A^{2} \nonumber \\ &\rightarrow-(\varepsilon_{d}-\mu)\frac{\dot{\varepsilon}_{d}^{2}}{2}f''(\varepsilon_{d},\mu)\,\frac{2}{\Gamma}\,,
\end{align}
which is identical to the quadratic contribution in Eq.\ \eqref{QClassical}. It is readily realized that the weak coupling limit of all the thermodynamic quantities in Secs.\ \ref{Equ.Thermodyn} and \ref{NonEqu.Thermodyn} are identical to the expressions Eqs.\ \eqref{ClassicalN}-\eqref{SdotClass} derived form the rate equation \eqref{MasterEq}.

\section{Conclusion}
\label{sec:Concl}

We have developed a consistent non-equilibrium quantum thermodynamics of the driven resonant level model where the effects of the driving are evaluated within the framework of non-equilibrium Green's functions and the gradient expansion. Our construction is consistent with the first and second laws of thermodynamics and with particle conservation. The problem of taking proper account of the strong system-bath coupling was circumvented by considering the extended resonant level -- the part of the overall system, or the 'world', that  is affected by local changes in the level energy. The method developed here of representing these equilibrium thermodynamic functions by quasistatic expectation values of operators and subsequently extending the model to finite level speed with the help of the non-equilibrium Green's functions formalism  can provide a guideline for future thermodynamic treatments of strongly coupled quantum systems. It should be kept in mind, however, that our model was restricted to a particular 
 kind of driving -- a time-dependent level energy -- and our calculations were done in the wide band limit. Extending our treatment to more general situations may require further theoretical considerations, with some difficulties already pointed out in Ref.\ \onlinecite{Esposito2014}. Another interesting problem is the inclusion of interactions of the dot electron with the electrons in the lead. Some thermodynamic properties have been been studied including these  interactions, in particular the specific heat and susceptibility in the context of Kondo systems \cite{Tsvelick1983} and the ohmic two-state system.\cite{Nghiem2016,Weiss1999} However, an inclusion of interactions into the full thermodynamic description of the driven level remains an open issue.

\section*{Acknowledgments}
We acknowledge fruitful discussions with L.\ Arrachea and M.\ Galperin. This work was supported in part by SFB 658 of the Deutsche Forschungsgemeinschaft, including a Mercator Professorship (A.N.). A.N.\ also acknowledges the support of the Israel Science Foundation.



%
\appendix
\onecolumngrid

\section{Calculating the density of states of the resonant level model}\label{DensityOfStates}

In the following we calculate the part of the density of states that
changes when the dot level is moved, which in turn determines the relevant thermodynamic
quantities of the extended resonant level. As in the main text, this derivation is presented without using the wide band limit to achieve better comparison with the recent work of Ref.\ \onlinecite{EspositoQThermo}. This density of states is given by the trace of the spectral function
$\rho \left(\varepsilon \right)=\sum _{n} \, A_{nn} (\varepsilon )$, 
where  $A_{nn} (\varepsilon )=-2\, \text{Im}\, G_{nn}^{R} (\varepsilon )$. In the basis of uncoupled dot (d) and lead free electron states (k) this gives
\begin{equation}
 \rho \left(\varepsilon \right)=A_{dd}(\varepsilon)+\sum_k A_{kk}(\varepsilon) \,.
\end{equation}
The spectral function of the dot electrons in presence of the coupling
takes the well known form 
\begin{equation}
A_{dd}(\varepsilon)=\frac{-2\, \text{Im}\,\Sigma^{R}(\varepsilon)}{\left(\varepsilon-\varepsilon_{d}-\, \text{Re}\,\Sigma^{R}(\varepsilon)\right)^{2}+\left(\, \text{Im}\,\Sigma^{R}(\varepsilon)\right)^{2}}
\end{equation}
where $\Sigma^{R}(\varepsilon)=\sum_{k}\left|V_{k}\right|^{2}g_{k}^{R}(\varepsilon)$
is the retarded self energy of the dot state due to its coupling to the leads and $g_{k}^{R}(\varepsilon)$
is the retarded Green's function of a free lead electron in state $k$. Due to the strong coupling of
the dot to the lead electrons, also the density of states of the surrounding
responds upon changes in the dot level. To calculate $A_{kk}$ we start from the Dyson equation
for $G_{kk}^{R}(\varepsilon)$ 
\begin{equation}\label{Gkk}
G_{kk}^{R}(\varepsilon)=g_{k}^{R}(\varepsilon)+\left(g_{k}^{R}(\varepsilon)\right)^{2}\abs{V_{k}}^{2}G_{dd}^{R}(\varepsilon)\,.
\end{equation}
Summing over $k$ and using $\Sigma^{R}(\varepsilon)=\sum_{k}\left|V_{k}\right|^{2}g_{k}^{R}(\varepsilon)$
we can write the second term on the right of Eq.\ \eqref{Gkk} in terms of the retarded
self energy, leading to the total density of states
\begin{equation}
\rho\left(\varepsilon\right)=A_{dd}(\varepsilon)\left(1-\frac{d}{d\varepsilon}\, \text{Re}\,\Sigma^{R}(\varepsilon)\right)+2\, \text{Re}\, G_{dd}^{R}(\varepsilon)\frac{d}{d\varepsilon}\, \text{Im}\,\Sigma^{R}(\varepsilon)+\nu(\varepsilon)\,,
\end{equation}
where $\nu(\varepsilon)=-2\sum_k \, \text{Im}\, g_{k}^{R}(\varepsilon)$ is the density of states of the free metal. This is Eq.\ \eqref{RhoFull} from the main text.

\section{Density of states in the wide band limit}\label{WBL}

In this work, we use the term wide band limit in the following sense: We consider a large bandwidth $2D$ in the lead with a constant product of coupling matrix element $\abs{V_k}^2$ and lead density of states $\nu(\varepsilon)$,
\begin{align}
\Gamma = 2\pi\nu(\varepsilon)\abs{V(\varepsilon)}^{2},
\end{align}
for energies $\varepsilon$ within the bandwidth of the lead. This leads to the retarded dot self energy
\begin{equation} \label{SigmaRWBLfull}
\Sigma ^{R} (\varepsilon)=\lim _{\eta \to 0} \sum _{k}\frac{\left|V_{k} \right|^{2} }{\varepsilon -\varepsilon _{k} +i\eta }  = \frac{\Gamma}{2\pi}\ln\abs{\frac{D+\varepsilon}{D-\varepsilon}}-i\frac{\Gamma}{2}\Theta\left(D-\abs{\varepsilon}\right)\,,
\end{equation} 
where $\Theta$ is the Heaviside function. 

For energies $\varepsilon \ll D$, we can approximate the real part as $\Sigma ^{R} (\varepsilon)\simeq 2\varepsilon/D$, which gives small corrections to the quasiparticle weight, the level energy $\varepsilon_d$, and the level width $\Gamma$. Neglecting this contribution in the limit $D\to\infty$, we find the approximation used in the bulk of the paper. 

Strictly speaking, this approximation leads to divergences. To see that these divergences do not lead to complications in our discussion of the thermodynamics, one needs to treat the wide band limit somewhat more carefully.  From Eq.\ \eqref{RhoFull} in the main text we obtain the  density of states in the wide band limit
\begin{align}
\rho\left(\varepsilon\right)=&\frac{\Gamma\Theta\left(D-\abs{\varepsilon}\right)}{\left(\varepsilon-\varepsilon_{d}-\frac{\Gamma}{2\pi}\ln\abs{\frac{D+\varepsilon}{D-\varepsilon}}\right)^{2}+\left(\frac{1}{2}\Gamma\Theta\left(D-\abs{\varepsilon}\right)\right)^{2}}\left(1-\frac{\Gamma}{2\pi}\frac{d}{d\varepsilon}\ln\abs{\frac{D+\varepsilon}{D-\varepsilon}}\right) \nonumber \\
&-\frac{\left(\varepsilon-\varepsilon_{d}-\frac{\Gamma}{2\pi}\ln\abs{\frac{D+\varepsilon}{D-\varepsilon}}\right)}{\left(\varepsilon-\varepsilon_{d}-\frac{\Gamma}{2\pi}\ln\abs{\frac{D+\varepsilon}{D-\varepsilon}}\right)^{2}+\left(\frac{1}{2}\Gamma\Theta\left(D-\abs{\varepsilon}\right)\right)^{2}}\frac{d}{d\varepsilon}\Gamma\Theta\left(D-\abs{\varepsilon}\right)+\nu(\varepsilon)\,.
\end{align} 
The large but finite bandwidth of the lead reduces the energy interval in which the density of states takes finite values to $\varepsilon\in[-D,D]$. The energy dependence of the self energy that arises from the finite bandwidth leads to additional contributions to the density of states of the extended resonant level (the full density of states $\rho$ minus the unperturbed density of states in the bath $\nu$) for energies close to the band edge $\varepsilon\sim \pm D$. To calculate the influence of these additional terms on the thermodynamic quantities, we consider their contribution to the quasistatic energy $E^{(0)} =\Omega +\mu N^{(0)}  + T S^{(0)}$  Eq. \eqref{GrindEQ__11_}, the quantity with the largest contribution from the band edge. The correction to the internal
energy $\delta E_1$ originating from the term $\propto\frac{d}{d\varepsilon}\Im\Sigma^{R}(\varepsilon)$
vanishes,
\begin{align}
\delta E_1 & =  -\int\frac{d\epsilon}{2\pi}\epsilon f(\epsilon)\frac{\left(\varepsilon-\varepsilon_{d}-\frac{\Gamma}{2\pi}\ln\abs{\frac{D+\epsilon}{D-\epsilon}}\right)}{\left(\varepsilon-\varepsilon_{d}-\frac{\Gamma}{2\pi}\ln\abs{\frac{D+\epsilon}{D-\epsilon}}\right)^{2}+\left(\frac{1}{2}\Gamma\Theta\left(D-\abs{\epsilon}\right)\right)^{2}}\frac{d}{d\varepsilon}\Gamma\Theta\left(D-\abs{\epsilon}\right)=0\,.
\end{align}
The correction $\delta E_2$ from the term $\propto\frac{d}{d\varepsilon}\Re\Sigma^{R}(\varepsilon)$ takes the form
\begin{align}
\delta E_2  =  & \int\frac{d\epsilon}{2\pi}\epsilon f(\epsilon)\frac{-\Gamma\Theta\left(D-\abs{\epsilon}\right)}{\left(\varepsilon-\varepsilon_{d}-\frac{\Gamma}{2\pi}\ln\abs{\frac{D+\epsilon}{D-\epsilon}}\right)^{2}+\left(\frac{1}{2}\Gamma\Theta\left(D-\abs{\epsilon}\right)\right)^{2}}\frac{\Gamma}{2\pi}\frac{2D}{D^{2}-\epsilon^{2}}\,,
\end{align}
where we used $\frac{d}{d\varepsilon}\ln\abs{\frac{D+\epsilon}{D-\epsilon}}= \frac{2D}{D^{2}-\epsilon^{2}}$. To estimate the correction from the band edge, consider the contribution from the upper edge $\varepsilon \sim D$. The divergence of $\frac{\Gamma}{2\pi}\ln\abs{\frac{D+\epsilon}{D-\epsilon}}$ dominates the denominator when
\begin{eqnarray}
D & \lesssim & \frac{\Gamma}{2\pi}\ln\abs{\frac{D+\epsilon}{D-\epsilon}}\\
2D\e^{-D/\Gamma} & \lesssim & D-\epsilon.
\end{eqnarray}
Hence we can separate the energy integral in $\delta E_2$ into the two parts,
\begin{align}
\delta E_2 & \simeq  \int^{D-2D\e^{-D/\Gamma}}\frac{d\epsilon}{2\pi}Df(D)\frac{-\Gamma}{D^{2}}\frac{\Gamma}{2\pi}\frac{2D}{2D(D-\epsilon)} \nonumber \\
 & +  \int_{D-2D\e^{-D/\Gamma}}^{D}\frac{d\epsilon}{2\pi}Df(D)\frac{-\Gamma}{\left(\frac{\Gamma}{2\pi}\ln\abs{\frac{D-\epsilon}{2D}}\right)^{2}}\frac{\Gamma}{2\pi}\frac{2D}{2D(D-\epsilon)}\,.
\end{align}
Estimating the integrals leads to 
\begin{align}
\delta E_2 & \simeq  -f(D)\frac{\Gamma}{\left(2\pi\right)^2}-\Gamma f(D)\,.
\end{align}
The contribution from the lower edge $\varepsilon \sim -D$ follows analogously and yields an analogous result with $f(D)$ replaced by $f(-D)$. Thus, the contribution to the density of states $\propto\frac{d}{d\varepsilon}\Re\Sigma^{R}(\varepsilon)$ gives a finite cutoff-dependent correction to the internal energy that does not vanish in the limit $D\rightarrow\infty$. 

However, the thermodynamics actually relates changes in the thermodynamic state functions, and not the state  functions themselves.  We can similarly consider how these changes are affected by starting with a finite bandwidth. To be specific, consider the change of the internal energy upon moving the dot level $\frac{d}{d\epsilon_{d}}\delta E$. By analogy with the above, the contribution $\propto\frac{d}{d\varepsilon}\Im\Sigma^{R}(\varepsilon)$ yields
\begin{align}
\frac{d}{d\epsilon_{d}}\delta E_2 & = \int\frac{d\epsilon}{2\pi}\epsilon f(\epsilon)\frac{-2\Gamma\Theta\left(D-\abs{\epsilon}\right)\left(\varepsilon-\varepsilon_{d}-\frac{\Gamma}{2\pi}\ln\abs{\frac{D+\epsilon}{D-\epsilon}}\right)}{\left(\left(\varepsilon-\varepsilon_{d}-\frac{\Gamma}{2\pi}\ln\abs{\frac{D+\epsilon}{D-\epsilon}}\right)^{2}+\left(\frac{1}{2}\Gamma\Theta\left(D-\abs{\epsilon}\right)\right)^{2}\right)^{2}}\frac{\Gamma}{2\pi}\frac{2D}{D^{2}-\epsilon^{2}}\\ 
 & \simeq \frac{-2 f(D)\Gamma}{\left(2\pi\right)^2 D} + \frac{4\pi\Gamma f(D)}{2D} \rightarrow0\text{ for }D\rightarrow\infty\,.
\end{align}
Hence for the \textit{changes of the thermodynamic quantities}, the corrections associated with the energy dependence of the self energy vanish in the limit $D\rightarrow\infty$ . The specific choice of the bandwidth $D$ merely sets the reference point from which the grand potential  $\Omega$ and the internal energy $E^{(0)}$ of the extended resonant level are being measured -- all changes of thermodynamic quantities and non-equilibrium corrections are converging to cutoff-independent results in the limit $D\rightarrow\infty$. This leads to the wide band limit expression for the density of states of the extended resonant level in the limit of large $D$
\begin{align}
 \rho_{\varepsilon_d}(\varepsilon)=\frac{\Gamma  }{\left(\varepsilon -\varepsilon _{d} \right)^{2} +\left(\Gamma /2\right)^{2} }
\end{align}
used in the main text, which leaves the dependence on $D$ that sets the reference point of the internal energy $E^{(0)}$ and the grand potential $\Omega$ implicit.

\section{Calculation of the non-equilibrium Green's functions of the resonant
level model}\label{NEGF}
Here we evaluate the necessary elements of the non-equilibrium Green functions for the driven resonant level model. The gradient expansion is utilized to take advantage of the model assumption that the driving speed is slow  relative to the electronic relaxation rates.

We start by deriving the form of the retarded dot
Green's function at finite speed $G_{dd}^{R}(t,t')=-i\Theta(t-t')\left\langle \left\{ d(t),d^{\dag}(t')\right\} \right\rangle $, where we omit the subscript in the following.
The equation of motion for the retarded Green's function can be written in the form
\begin{equation}
\delta(t-t')=\int dt_{1}G^{R}(t,t_{1})\left[i\partial_{t_1}\delta(t_{1}-t')-\varepsilon_{d}(t_{1})\delta(t_{1}-t')-\Sigma^{R}(t_{1}-t')\right]\label{eq:EOM GR}
\end{equation}
with the retarded self energy $\Sigma^{R}(t,t')=\sum_{k}\left|V_{k}\right|^{2}g_{k}^{R}(t,t')$.
To perform an adiabatic expansion it is beneficial to switch to a
description in terms of Wigner transforms 
\begin{equation}
G\left(\varepsilon,t\right)=\int d\tau \, G\left(t_{1},t_{2}\right)e^{i\varepsilon\tau}\,,
\end{equation}
where $t=\frac{t_{1}+t_{2}}{2}$ and $\tau=t_{1}-t_{2}$ and the corresponding
inverse transform. Using that the Wigner transform of a convolution can be written
like
\begin{equation}
\int C\left(t_{1},t_{3}\right)D\left(t_{3},t_{2}\right)dt_{3}=\int\frac{d\varepsilon}{2\pi}e^{-i\varepsilon\tau}C\left(\varepsilon,t\right)*D\left(\varepsilon,t\right)
\end{equation}
with $C\left(\varepsilon,t\right)*D\left(\varepsilon,t\right)=C\left(\varepsilon,t\right)\exp\left[\frac{i}{2}\left(\stackrel{\leftarrow}{\partial}_{\varepsilon}\vec{\partial}_{t}-\stackrel{\leftarrow}{\partial}_{t}\vec{\partial}_{\varepsilon}\right)\right]D\left(\varepsilon,t\right)$
we can take the Wigner transform of Eq.\ \eqref{eq:EOM GR} and expand
the exponential up to first order to obtain
\begin{eqnarray}
1 & = & G^{R}(\varepsilon,t)\left[\varepsilon-\varepsilon_{d}(t)+\frac{1}{2}i\Gamma\right]+\frac{i}{2}\left[\partial_\varepsilon G^{R}(\varepsilon,t)\left[-\dot{\varepsilon}_{d}(t)\right]-\partial_{t}G^{R}(\varepsilon,t)\right]\,,
\end{eqnarray}
where we used the wide band limit $\Sigma^{R}=-\frac{1}{2}i\Gamma$.
Thus the retarded Green's function of the dot electrons is, up
to first order in the velocity, given by the frozen form $G^{R}(\varepsilon,t)=(\varepsilon-\varepsilon_{d}(t)+i\frac{\Gamma}{2})^{-1}$.
An analogous calculation gives for the advanced Green's function $G^{A}(t,t')=i\Theta(t'-t)\left\langle \left\{ d(t),d^{\dag}(t')\right\} \right\rangle $
the Wigner transform $G^{A}(\varepsilon,t)=(\varepsilon-\varepsilon_{d}(t)-i\frac{\Gamma}{2})^{-1}$.

The lesser Green's function of the dot electrons $G^{<}(t,t')=i\left\langle d^{\dag}(t')d(t)\right\rangle $
can be calculated via the Langreth rule given the lesser component
of the self energy $\Sigma^{<}$,\cite{JauhoTime-DependentTransp}
\begin{equation}
G^{<}(t,t')=\int dt_{1}dt_{2}G^{R}(t,t_{1})\Sigma^{<}\left(t_{1},t_{2}\right)G^{A}(t_{2},t')
\end{equation}
Note that the Green's function can be alternatively calculated in the partition-free approach to quantum transport, which assumes equilibration of the entire 
	system in presence of the coupling as boundary conditions for the Green's functions.\cite{Stefanucci2004} 
Taking the Wigner transform of this convolution and expanding up to
first order in the velocity we obtain in the different orders
\begin{eqnarray}
G^{<(0)}(\varepsilon,t) & = & G^{R}\Sigma^{<}G^{A}\,, \\
G^{<(1)}(\varepsilon,t) & = & \frac{i}{2}\left(\partial_{\varepsilon}G^{R}\partial_{t}\Sigma^{<}-\partial_{t}G^{R}\partial_{\varepsilon}\Sigma^{<}\right)G^{A}+\frac{i}{2}\left[\partial_{\varepsilon}\left(G^{R}\Sigma^{<}\right)\partial_{t}G^{A}-\partial_{t}\left(G^{R}\Sigma^{<}\right)\partial_{\varepsilon}G^{A}\right]\,.
\end{eqnarray}
Using $\partial_{t}G^{R/A}=-\dot{\varepsilon}_{d}\partial_{\varepsilon}G^{R/A}$, $\Sigma^{<}(\varepsilon)=if\left(\varepsilon\right)\Gamma$ and $\partial_{\varepsilon}G^{R}\, G^{A}-G^{R}\partial_{\varepsilon}G^{A}=i\frac{A^2}{\Gamma}$ we obtain
\begin{equation}\label{Glesser}
G^{<}(\varepsilon,t)=iA\, f-i\frac{\dot{\varepsilon}_{d}}{2}\partial_{\varepsilon}f\, A^{2}\,.
\end{equation}

\section{An alternative derivation of the non-equilibrium Green's functions in terms of a quantum kinetic equation}\label{QBoltzmann}

 Here we present an alternative derivation of the non-equilibrium properties of the dot electrons. Instead of deriving the lesser component of the non-equilibrium Green's function using the Langreth rule in the Keldysh integral formulation, one can equivalently derive the non-equilibrium occupation of the level using the a quantum kinetic (Kadanoff-Baym or Quantum Boltzmann) equation in first order gradient approximation,\cite{haug1996quantum} as it is done in Ref.\ \onlinecite{EspositoQThermo} and \onlinecite{Kita2010}. For the description of a single electronic level in contact to leads these approaches are equivalent and we explicitly show both here to clarify the connection of our work to Ref.\ \onlinecite{EspositoQThermo}. For a single electronic level, the retarded Green's function of the dot electrons takes the frozen form $G^{R}(\varepsilon,t)=(\varepsilon-\varepsilon_{d}(t)+i\frac{\Gamma}{2})^{-1}$ 
when considering the gradient expansion of the Dyson equation up to second order. Thereby also the form of the spectral function $A(\varepsilon )=-2\, \text{Im}\, G^{R} (\varepsilon )$  is set and all effects of the level speed up to linear order can be cast in a non-equilibrium distribution function $\phi$, related to the lesser Green's function via $G^{<} =iA\phi $. The non-equilibrium distribution function of the dot electrons in contact to one lead satisfies the equation of motion \cite{Kita2010}
\begin{equation} \label{QBoltzmannEq}
\{ G_{0}^{-1}-\, \text{Re}\, \Sigma ^{R}  ,\; A\phi \} -\{ \Gamma f, \, \text{Re}\, G^{R} \} =A\Gamma (f-\phi)\,,  
\end{equation}
where $ \left\{C,\; D\right\} = \partial_\varepsilon C \, \partial_t D  -\partial_t C \, \partial_\varepsilon D    $ is the Poisson bracket and $G_{0}^{-1}=\varepsilon-\varepsilon_d(t)$. \footnote{Eq.\ \eqref{QBoltzmannEq} differs from Eq.\ (4.19) in Ref.\ \onlinecite{Kita2010}, used in Ref.\ \onlinecite{EspositoQThermo}, in the second Poisson bracket on the left, as our expression involves $f$ rather than $\phi$. We believe that our form is correct, but in any case both forms are equivalent up to the first order in velocity considered here and both lead to the same solution for $\phi$  Eq.\ \eqref{PhiQBoltzmann}.} Using the wide band limit we solve this equation for $\phi$ consistently up to linear in the velocity to obtain
\begin{align}\label{PhiQBoltzmann}
 \phi&=f-\dot{\varepsilon}_d \partial_\varepsilon f \left(\frac{1}{\Gamma}+\partial_\varepsilon \, \text{Re}\, G^R \right)\nonumber \\
 &=f-\frac{\dot{\varepsilon}_d}{2} \partial_\varepsilon f A\,,
\end{align}
which is identical to the solution above obtained via the Langreth rule for the lesser component of the Green's function Eq.\  \eqref{Glesser}.

\section{Calculation of the internal energy}\label{IntEnergyAp}
As mentioned in the main text, the internal energy of the extended resonant level model can be, at different orders $i$, represented as expectation value of the Hamiltonian of the effective system $H_D + \frac{1}{2} H_V$ 
\begin{equation}
E^{(i)}=\braket{H_{D}}^{(i)}+\frac{1}{2}\braket{H_{V}}^{(i)}  \,.
\end{equation}
To calculate $\braket{H_{V}}$ we write 
\begin{align}
\braket{H_{V}} = & \sum_{k}\left(V_{k}\braket{d^{\dagger}c_{k}}+V_{k}^{*}\braket{c_{k}^{\dagger}d}\right)\\
  = & 2\sum_{k}\, \text{Im}\,\left(V_{k}^{*}G_{d,k}^{<}(t,t)\right)\,,
\end{align}
with $G_{d,k}^{<}(t,t')= i\left\langle c_{k}^{\dag}(t')d(t)\right\rangle $ and
where we used $G_{d,k}^{<}(t,t)=-\left(G_{k,d}^{<}(t,t)\right)^{*}$.
The equation of motion for the mixed Green's function $G_{d,k}^{<} $ and analytical
continuation from the Keldysh contour to the lesser component leads
to \cite{JauhoTime-DependentTransp}
\begin{align}
\braket{H_{V}}&=2\sum_{k}\, \text{Im}\,\left(\int dt'\abs{V_{k}}^{2}\left[G^{R}(t,t')g_{k}^{<}(t',t)+G^{<}(t,t')g_{k}^{A}(t',t)\right]\right) \nonumber \\
&=2\, \text{Im}\,\left(\int dt'\left[G^{R}(t,t')\Sigma^{<}(t',t)+G^{<}(t,t')\Sigma^{A}(t',t)\right]\right)\,. 
\end{align}
Moving to the Wigner transform we obtain
\begin{equation}
\braket{H_{V}}=2\, \text{Im}\,\left(\int\frac{d\varepsilon}{2\pi}\left[G^{R}(\varepsilon,t)*\Sigma^{<}(\varepsilon)+G^{<}(\varepsilon,t)*\Sigma^{A}\right]\right)\,.
\end{equation}
Note that $G^{<}(\varepsilon,t)*\Sigma^{A}=G^{<}(\varepsilon,t)\,\frac{1}{2}i\Gamma$ does not contribute, since it is purely real. 
This leads up to linear order in the velocity to 
\begin{equation}\label{CouplingHLinear}
\braket{H_{V}}=2\, \text{Im}\,\left(\int\frac{d\varepsilon}{2\pi}\left[G^{R}(\varepsilon,t)if(\varepsilon)\Gamma-\frac{i}{2}\partial_{t}G^{R}(\varepsilon,t)i\partial_{\varepsilon}f(\varepsilon)\Gamma\right]\right)\,.
\end{equation}

From the fact that $G^R$ does not have a correction linear in the velocity it follows that the first term on the right contributes only in zero order, and yields the quasistatic coupling energy $\braket{H_{V}}^{(0)}$
\begin{eqnarray}\label{H_V_0}
\braket{H_{V}}^{(0)} & = & \int\frac{d\varepsilon}{2\pi}2f(\varepsilon)\Gamma\, \text{Re}\, G^{R}(\varepsilon)\\
 & = & 2\int\frac{d\varepsilon}{2\pi}f(\varepsilon)(\varepsilon-\varepsilon_{d})A\,,
\end{eqnarray}
which leads, using the result for $\braket{H_{D}}^{(0)}=\varepsilon_d\braket{d^{\dag}d}^{(0)}$ from Eq.\ \eqref{Glesser}, to the quasistatic internal energy of the extended
resonant level given in the main text Eq.\ \eqref{GrindEQ__11_}
\begin{equation}
E^{(0)}=\braket{H_{D}}^{(0)}+\frac{1}{2}\braket{H_{V}}^{(0)}=\int\frac{d\varepsilon}{2\pi}\varepsilon fA\,.
\end{equation}
The first order correction to  the coupling energy is obtained from the second term on the right of Eq.\ \eqref{CouplingHLinear}  and takes the form
\begin{eqnarray}\label{H_V_1}
\braket{H_{V}}^{(1)} & = & \int\frac{d\varepsilon}{2\pi}\partial_{\varepsilon}f\Gamma\, \text{Im}\, \partial_{t}G^{R}(\varepsilon)\\
& = & \frac{\dot{\varepsilon}_{d}}{2}\int\frac{d\varepsilon}{2\pi}\partial_{\varepsilon}f\Gamma \partial_{\varepsilon}A\,.
\end{eqnarray}
With $\braket{H_{D}}^{(1)}=\varepsilon_d\braket{d^{\dag}d}^{(1)}$ from Eq.\ \eqref{Glesser} we obtain the correction to the internal energy Eq.\ \eqref{IntEnergy1st} from the main text
\begin{align}
E^{(1)}=\braket{H_{D}}^{(1)}+\frac{1}{2}\braket{H_{V}}^{(1)}&=\dot\varepsilon_{d}\int\frac{d\varepsilon}{2\pi}\left(-\frac{\varepsilon_d}{2} \partial_{\varepsilon}fA^2+\frac{1}{4}\Gamma\partial_{\varepsilon}f \partial_{\varepsilon}A\right) \\
&=\frac{-\dot{\varepsilon}_{d}}{2} \int\frac{d\varepsilon}{2\pi}\varepsilon\partial_{\varepsilon}f A^2  \,,
\end{align}
where we used $\partial_{\varepsilon}A=\frac{-2(\varepsilon-\varepsilon_d)}{\Gamma }A^2$.
Taking the time derivative of this correction leads to the second order contribution to the internal energy change per unit time
\begin{align}
\frac{d}{dt}E^{(1)}=\dot{E}^{\left(2\right)} & = \frac{\dot\varepsilon_{d}^2}{2}\int\frac{d\varepsilon}{2\pi} \varepsilon \partial_{\varepsilon} f \partial_{\varepsilon}A^2 \\
& =  \frac{\dot\varepsilon_{d}^2}{2}\int\frac{d\varepsilon}{2\pi}\left(-\partial_{\varepsilon}fA^2-\varepsilon\partial^2_{\varepsilon}f\,A^2\right)\,,
\end{align}
where we integrated by parts. Note again that throughout the entire paper we assume a linear motion of the dot level $\ddot{\varepsilon}_d=0$. With the corresponding expressions given in the main text Eq.\ \eqref{GrindEQ__22_}, \eqref{GrindEQ__26_} and \eqref{GrindEQ__20_} it can be seen that the derived corrections satisfy the first law   $\dot{E}^{\left(2\right)} =\dot{W}^{\left(2\right)} +\dot{Q}^{\left(2\right)} +\mu \, \dot{N} ^{\left(2\right)} $.

Note however that even though the symmetric splitting into effective system and bath gives a correct representation of the  $\varepsilon_d$-dependent part of the internal energy (the internal energy of the extended resonant level model), it does not mean that $\braket{H_B}$ has no $\varepsilon_d$-dependent part. This can be seen explicitly by calculating the $\varepsilon_d$-dependent part of the lead Hamiltonian, which we call $\braket{H_B}_{\varepsilon_d}$ in the following, via a scaled version of the grand potential of the extended resonant level $\Omega$ Eq.\ \eqref{GrindEQ__9_} (the $\varepsilon_d$-dependent part of the grand potential).
We use the scaled Hamiltonian 
\begin{equation}
H_{\lambda}=H_{D}+\lambda H_{B}+H_{V}
\end{equation}
to calculate  $\braket{H_B}_{\varepsilon_d}$ from the associated scaled grand potential $\Omega_\lambda$
\begin{equation}
 \braket{H_B}_{\varepsilon_d}=\frac{\partial\Omega_\lambda}{\partial\lambda}\big{\vert}_{\lambda=1},
\end{equation}
evaluated at $\lambda=1$. The scaled lead Hamiltonian changes the density of states of the bath electrons $\nu_{\lambda}(\varepsilon)=\nu(\varepsilon)/\lambda$ and the scaled spectral function of the dot electrons $A_{\lambda}$ reads
\begin{equation}
A_{\lambda}=\frac{\Gamma}{\left(\varepsilon-\varepsilon_{d}\right)^{2}+\left(\frac{\Gamma}{2\lambda}\right)^{2}}\,.
\end{equation}
This sets the form of the scaled grand potential $\Omega_\lambda$ from which we obtain 
\begin{align}
\braket{H_B}_{\varepsilon_d} & =  \frac{-1}{\beta}\frac{\partial}{\partial\lambda} \int \frac{d\varepsilon }{2\pi } \frac{\Gamma}{\left(\varepsilon-\varepsilon_{d}\right)^{2}+\left(\frac{\Gamma}{2\lambda}\right)^{2}} \ln \left(1+e^{-\beta \left(\varepsilon -\mu \right)} \right) \\
&=-\frac{1}{\beta}\left(-\frac{\Gamma}{\lambda^{2}}\right)\int\frac{d\varepsilon}{2\pi}\underbrace{\frac{\left(\varepsilon-\varepsilon_{d}\right)^{2}-\left(\frac{\Gamma}{2\lambda}\right)^{2}}{\left[\left(\varepsilon-\varepsilon_{d}\right)^{2}+\left(\frac{\Gamma}{2\lambda}\right)^{2}\right]^{2}}}_{-\partial_\varepsilon ReG^{R}\left(\varepsilon\right)}\ln\left(1+e^{-\beta\left(\varepsilon-\mu\right)}\right)\quad \lambda \rightarrow 1	\\
 &=\frac{1}{\beta}\int\frac{d\varepsilon}{2\pi}\Gamma \, \text{Re}\, G^{R}\left(\varepsilon\right)\, \partial_\varepsilon\ln \left(1+e^{-\beta \left(\varepsilon -\mu \right)} \right)	\\
&=-\int\frac{d\varepsilon}{2\pi}\left(\varepsilon-\varepsilon_{d}\right)\, Af\left(\varepsilon\right) \,.
\end{align}
Note that an analogous calculation for $H_D$ and $H_V$ reproduces the direct expectation values $\braket{H_V}_{\varepsilon_d}=\braket{H_V}^{(0)}$ Eq.\ \eqref{H_V_0} and  $\braket{H_D}_{\varepsilon_d}=\varepsilon_d\braket{d^{\dag}d}^{(0)}$ from Eq.\ \eqref{Glesser}.
Thus the $\varepsilon_d$-dependent part of all three Hamiltonian reproduces the adiabatic internal energy of the extended resonant level from above
\begin{equation}
\braket{H_{D}}^{(0)}+\braket{H_{V}}^{(0)}+\braket{H_B}_{\varepsilon_d}^{(0)}=\int\frac{d\varepsilon}{2\pi}\varepsilon fA\,,
\end{equation}
while the sum $\braket{H_{V}}^{(0)}+\braket{H_B}_{\varepsilon_d}^{(0)}$ gives the "half splitting" contribution $\frac{1}{2}\braket{H_{V}}^{(0)}$.

\section{Calculation of the energy fluxes}\label{HeatFluxAp}

Using the results of App.\ \ref{NEGF} we can now calculate the different energy fluxes contributing to the heat current at different orders from the non-equilibrium Green's functions formalism. Since the energy fluxes $W_\alpha=i\braket{\left[H_{\rm tot},H_{\alpha}\right]}$ between the different parts of the system $\alpha$ must satisfy
\begin{equation}
 W_{B}+W_{V}+W_{D}=0\,,
\end{equation}
and because the energy change of the \textit{total} system is given by the power provided by the external driving $\dot{E}_{\rm tot}=\braket{\frac{\partial H_{d}}{\partial t}}$, there are in principle two ways of calculating the energy flow into the effective bath (needed for the evaluation of the heat flow at different orders):
\begin{align}
\dot{Q}&=-\left(\frac{1}{2} W_V-W_B\right) -\mu\dot{N} \quad \text{or}\\
 \dot{Q}&=W_D +\frac{1}{2} W_V -\mu\dot{N}\,.
\end{align}
We present the calculation via the energy flux leaving the effective system $W_D +\frac{1}{2} W_V $, since it takes a simpler form in the non-equilibrium Green's functions formalism. 
Note however that a calculation via $W_B$ is also possible and leads to the same result.

We calculate the heat flux via
\begin{align}
 \dot{Q}&=W_D +\frac{1}{2} W_V -\mu\dot{N}\nonumber \\
 &=\varepsilon_d\dot{N}+\frac{1}{2}\frac{d}{dt}\braket{H_{V}}-\mu\dot{N}\,.
\end{align}
This leads with $\dot{N}^{(1)}$ Eq.\ \eqref{GrindEQ__15_} and $\braket{H_{V}}^{(0)}$ Eq.\ \eqref{H_V_0} to the quasistatic heat current linear in the velocity 
\begin{align}
 \dot{Q}^{(1)}&= \varepsilon_d\dot{N}^{(1)}+\frac{1}{2}\frac{d}{dt}\braket{H_{V}}^{(0)}-\mu\dot{N}^{(1)} \nonumber \\
 &=\varepsilon_d\dot{\varepsilon}_{d}\int\frac{d\varepsilon}{2\pi}A\,\partial_\varepsilon f-\dot{\varepsilon}_{d}\int\frac{d\varepsilon}{2\pi}f\Gamma\partial_{\varepsilon}\, \text{Re}\, G^{R}-\mu\int \frac{d\varepsilon }{2\pi } A\,\partial_\varepsilon f\left(\varepsilon \right) \nonumber \\
 &= \dot{\varepsilon }_{d} \int \frac{d\varepsilon }{2\pi } \left(\varepsilon -\mu \right)A\,\partial_\varepsilon f\left(\varepsilon \right)  \,, 
\end{align}
where we used $\Gamma\, \text{Re}\, G^{R}=(\varepsilon-\varepsilon_{d})\, A$ and integrated by parts. Therefore the calculation of the first order heat current via the energy flux into the effective bath reproduces the adiabatic heat current Eq.\ \eqref{GrindEQ__14_} from the main text. 
To calculate the non-equilibrium correction we use  $\dot{N}^{(2)} $, Eq.\ \eqref{GrindEQ__20_}, and $\braket{H_{V}}^{(1)}$, Eq.\ \eqref{H_V_1}, and obtain
\begin{align}
 \dot{Q}^{(2)}&= \varepsilon_d\dot{N}^{(2)}+\frac{1}{2}\frac{d}{dt}\braket{H_{V}}^{(1)}-\mu\dot{N}^{(2)} \nonumber \\
 &=-\varepsilon_d\int\frac{d\varepsilon}{2\pi}\frac{\dot{\varepsilon}_{d}^{2}}{2}\,\partial^2_\varepsilon f\, A^{2} -\frac{\dot{\varepsilon}_{d}^{2}}{4}\int\frac{d\varepsilon}{2\pi}\Gamma\partial_{\varepsilon}f\partial_{\varepsilon}^{2}A-\mu\int\frac{d\varepsilon}{2\pi}\frac{\dot{\varepsilon}_{d}^{2}}{2}\,\partial^2_\varepsilon f\, A^{2} \nonumber \\
&=-\int\frac{d\varepsilon}{2\pi}(\varepsilon-\mu)\frac{\dot{\varepsilon}_{d}^{2}}{2}\,\partial^2_\varepsilon f\, A^{2}\,,\label{Q2NEGF} 
\end{align}
where we integrated by parts and used $ \Gamma\partial_{\varepsilon}A=-2(\varepsilon-\varepsilon_{d})A^{2}$. This is the form of the non-equilibrium correction to the heat current given in the main text Eq.\ \eqref{GrindEQ__26_}.

\section{Particle conservation of the finite speed current}\label{particle conservation}

In the following we show that the correction $\dot{N}^{(2)}=\frac{d}{dt}N^{(1)}$
to the quasistatic current is obeying particle conservation
upon moving on a path between two states with a well defined particle
number. We need to show that

\begin{equation}
\Delta N^{(2)}=\int_{t_{1}}^{t_{2}}dt\dot{N}^{(2)}=0\,,
\end{equation}
with $\varepsilon_{d}(t_{1})$ well below and $\varepsilon_{d}(t_{2})$
well above $\mu$. Assuming a constant velocity $\dot{\varepsilon}_{d}$
we obtain 
\begin{eqnarray}
\Delta N^{(2)} & = & -\int_{t_{1}}^{t_{2}}dt\int\frac{d\varepsilon}{2\pi}\frac{\dot{\varepsilon}_{d}^{2}}{2}\partial_{\varepsilon}^{2}f\, A^{2}\nonumber \\
 & = & \int_{\varepsilon_{1}}^{\varepsilon_{2}}d\varepsilon_{d}\frac{\dot{\varepsilon}_{d}}{2}\int\frac{d\varepsilon}{2\pi}\,\partial_\varepsilon f \partial_\varepsilon A^{2},
\end{eqnarray}
where we did an integration by parts in the second line. Now we use
that $A$ is a function of $\varepsilon-\varepsilon_{d}$ and therefore
$\,\partial_\varepsilon A=-\,\partial_{\varepsilon_d} A$
to obtain 
\begin{align*}
\Delta N^{(2)}  = & -\frac{\dot{\varepsilon}_{d}}{2}\int\frac{d\varepsilon}{2\pi}\,\partial_\varepsilon f\int_{\varepsilon_{1}}^{\varepsilon_{2}}d\varepsilon_{d}\frac{\partial A^{2}}{\partial\varepsilon_{d}}\\
  = & -\frac{\dot{\varepsilon}_{d}}{2}\int\frac{d\varepsilon}{2\pi}\,\partial_\varepsilon fA^{2}\vert_{\varepsilon-\varepsilon_{1}}^{\varepsilon-\varepsilon_{2}}\\
  = & 0\,,
\end{align*}
where we used that the derivative of the fermi distribution $\,\partial_\varepsilon f$
restricts the $\varepsilon$-interval in which the integrand is non-zero
to a finite range $\sim k_{B}T$ around $\mu$. As long as $\varepsilon_{1}$
is well below and $\varepsilon_{2}$ is well above it, $A^{2}(\varepsilon,\varepsilon_{1/2})$
is zero everywhere, where $\,\partial_\varepsilon f$ is
nonzero, from which follows the last line.

\section{Energy conservation of the corrections to heat current and the extra
work}\label{HeatWorkCompens}

In the following we show that all the extra work paid for moving the
level at finite speed is given as extra heat to the leads 

\begin{equation}
\int_{t_{1}}^{t_{2}}dt\dot{W}^{(2)}=-\int_{t_{1}}^{t_{2}}dt\dot{Q}^{(2)}\,,
\end{equation}
where again $\varepsilon_{d}(t_{1})$ is well below and $\varepsilon_{d}(t_{2})$ is 
well above $\mu$. With analogous steps as above we obtain assuming
a constant level speed
\begin{align*}
\int_{t_{1}}^{t_{2}}dt\dot{W}^{(2)}  = & -\int_{t_{1}}^{t_{2}}dt\dot{Q}^{(2)}\\
-\int_{t_{1}}^{t_{2}}dt\,\int\frac{d\varepsilon}{2\pi}\,\frac{\dot{\varepsilon}_{d}^{2}}{2}\partial_{\varepsilon}f\, A^{2}  = & \int_{t_{1}}^{t_{2}}dt\int\frac{d\varepsilon}{2\pi}\varepsilon\left(\frac{\dot{\varepsilon}_{d}^{2}}{2}\partial_{\varepsilon}^{2}f\, A^{2}\right)\\
\dot{\varepsilon}_{d}\,\int_{\varepsilon_{1}}^{\varepsilon_{2}}d\varepsilon_{d}\,\int\frac{d\varepsilon}{2\pi}\,\varepsilon\partial_{\varepsilon}\left(\partial_{\varepsilon}f\, A^{2}\right)  = & \dot{\varepsilon}_{d}\,\int_{\varepsilon_{1}}^{\varepsilon_{2}}d\varepsilon_{d}\,\int\frac{d\varepsilon}{2\pi}\varepsilon\partial_{\varepsilon}^{2}f\, A^{2}\\
\int_{\varepsilon_{1}}^{\varepsilon_{2}}d\varepsilon_{d}\,\int\frac{d\varepsilon}{2\pi}\,\varepsilon\partial_{\varepsilon}f\,\partial_{\varepsilon}A^{2}  = & 0\\
-\int\frac{d\varepsilon}{2\pi}\,\varepsilon\partial_{\varepsilon}f\,\int_{\varepsilon_{1}}^{\varepsilon_{2}}d\varepsilon_{d}\partial_{\varepsilon_{d}}A^{2}  = & 0\\
-\int\frac{d\varepsilon}{2\pi}\,\varepsilon\partial_{\varepsilon}f\, A^{2}\vert_{\varepsilon-\varepsilon_{1}}^{\varepsilon-\varepsilon_{2}}  = & 0\,,
\end{align*}
where the last equality is fulfilled due to the finite range where
$\partial_{\varepsilon}f$ is non-zero, as above.


%
%
%
%
%

%


\begin{thebibliography}{26}%
\makeatletter
\providecommand \@ifxundefined [1]{%
 \@ifx{#1\undefined}
}%
\providecommand \@ifnum [1]{%
 \ifnum #1\expandafter \@firstoftwo
 \else \expandafter \@secondoftwo
 \fi
}%
\providecommand \@ifx [1]{%
 \ifx #1\expandafter \@firstoftwo
 \else \expandafter \@secondoftwo
 \fi
}%
\providecommand \natexlab [1]{#1}%
\providecommand \enquote  [1]{``#1''}%
\providecommand \bibnamefont  [1]{#1}%
\providecommand \bibfnamefont [1]{#1}%
\providecommand \citenamefont [1]{#1}%
\providecommand \href@noop [0]{\@secondoftwo}%
\providecommand \href [0]{\begingroup \@sanitize@url \@href}%
\providecommand \@href[1]{\@@startlink{#1}\@@href}%
\providecommand \@@href[1]{\endgroup#1\@@endlink}%
\providecommand \@sanitize@url [0]{\catcode `\\12\catcode `\$12\catcode
  `\&12\catcode `\#12\catcode `\^12\catcode `\_12\catcode `\%12\relax}%
\providecommand \@@startlink[1]{}%
\providecommand \@@endlink[0]{}%
\providecommand \url  [0]{\begingroup\@sanitize@url \@url }%
\providecommand \@url [1]{\endgroup\@href {#1}{\urlprefix }}%
\providecommand \urlprefix  [0]{URL }%
\providecommand \Eprint [0]{\href }%
\providecommand \doibase [0]{http://dx.doi.org/}%
\providecommand \selectlanguage [0]{\@gobble}%
\providecommand \bibinfo  [0]{\@secondoftwo}%
\providecommand \bibfield  [0]{\@secondoftwo}%
\providecommand \translation [1]{[#1]}%
\providecommand \BibitemOpen [0]{}%
\providecommand \bibitemStop [0]{}%
\providecommand \bibitemNoStop [0]{.\EOS\space}%
\providecommand \EOS [0]{\spacefactor3000\relax}%
\providecommand \BibitemShut  [1]{\csname bibitem#1\endcsname}%
\let\auto@bib@innerbib\@empty
\bibitem [{\citenamefont {Kay}\ and\ \citenamefont {Leigh}(2015)}]{Kay2015}%
  \BibitemOpen
  \bibfield  {author} {\bibinfo {author} {\bibfnamefont {E.~R.}\ \bibnamefont
  {Kay}}\ and\ \bibinfo {author} {\bibfnamefont {D.~A.}\ \bibnamefont
  {Leigh}},\ }\href@noop {} {\bibfield  {journal} {\bibinfo  {journal} {Angew.
  Chemie Int. Ed.}\ }\textbf {\bibinfo {volume} {54}},\ \bibinfo {pages}
  {10080} (\bibinfo {year} {2015})}\BibitemShut {NoStop}%
\bibitem [{Note1()}]{Note1}%
  \BibitemOpen
  \bibinfo {note} {These issues are not limited to the quantum regime. However,
  the treatment of ultrasmall systems frequently requires quantum
  considerations.}\BibitemShut {Stop}%
\bibitem [{\citenamefont {Ludovico}\ \emph {et~al.}(2014)\citenamefont
  {Ludovico}, \citenamefont {Lim}, \citenamefont {Moskalets}, \citenamefont
  {Arrachea},\ and\ \citenamefont {S\'anchez}}]{LilianaDynamical}%
  \BibitemOpen
  \bibfield  {author} {\bibinfo {author} {\bibfnamefont {M.~F.}\ \bibnamefont
  {Ludovico}}, \bibinfo {author} {\bibfnamefont {J.~S.}\ \bibnamefont {Lim}},
  \bibinfo {author} {\bibfnamefont {M.}~\bibnamefont {Moskalets}}, \bibinfo
  {author} {\bibfnamefont {L.}~\bibnamefont {Arrachea}}, \ and\ \bibinfo
  {author} {\bibfnamefont {D.}~\bibnamefont {S\'anchez}},\ }\href@noop {}
  {\bibfield  {journal} {\bibinfo  {journal} {Phys. Rev. B}\ }\textbf {\bibinfo
  {volume} {89}},\ \bibinfo {pages} {161306} (\bibinfo {year}
  {2014})}\BibitemShut {NoStop}%
\bibitem [{\citenamefont {Esposito}\ \emph
  {et~al.}(2015{\natexlab{a}})\citenamefont {Esposito}, \citenamefont {Ochoa},\
  and\ \citenamefont {Galperin}}]{EspositoQThermo}%
  \BibitemOpen
  \bibfield  {author} {\bibinfo {author} {\bibfnamefont {M.}~\bibnamefont
  {Esposito}}, \bibinfo {author} {\bibfnamefont {M.~A.}\ \bibnamefont {Ochoa}},
  \ and\ \bibinfo {author} {\bibfnamefont {M.}~\bibnamefont {Galperin}},\
  }\href@noop {} {\bibfield  {journal} {\bibinfo  {journal} {Phys. Rev. Lett.}\
  }\textbf {\bibinfo {volume} {114}},\ \bibinfo {pages} {080602} (\bibinfo
  {year} {2015}{\natexlab{a}})}\BibitemShut {NoStop}%
\bibitem [{\citenamefont {Allahverdyan}\ and\ \citenamefont
  {Nieuwenhuizen}(2000)}]{Allahverdyan2000a}%
  \BibitemOpen
  \bibfield  {author} {\bibinfo {author} {\bibfnamefont {A.~E.}\ \bibnamefont
  {Allahverdyan}}\ and\ \bibinfo {author} {\bibfnamefont {T.~M.}\ \bibnamefont
  {Nieuwenhuizen}},\ }\href@noop {} {\bibfield  {journal} {\bibinfo  {journal}
  {Phys. Rev. Lett.}\ }\textbf {\bibinfo {volume} {85}},\ \bibinfo {pages}
  {1799} (\bibinfo {year} {2000})}\BibitemShut {NoStop}%
\bibitem [{\citenamefont {H\"{a}nggi}\ \emph {et~al.}(2008)\citenamefont
  {H\"{a}nggi}, \citenamefont {Ingold},\ and\ \citenamefont
  {Talkner}}]{Hanggi2008}%
  \BibitemOpen
  \bibfield  {author} {\bibinfo {author} {\bibfnamefont {P.}~\bibnamefont
  {H\"{a}nggi}}, \bibinfo {author} {\bibfnamefont {G.-L.}\ \bibnamefont
  {Ingold}}, \ and\ \bibinfo {author} {\bibfnamefont {P.}~\bibnamefont
  {Talkner}},\ }\href@noop {} {\bibfield  {journal} {\bibinfo  {journal} {New
  J. Phys.}\ }\textbf {\bibinfo {volume} {10}},\ \bibinfo {pages} {115008}
  (\bibinfo {year} {2008})}\BibitemShut {NoStop}%
\bibitem [{\citenamefont {Hilt}\ and\ \citenamefont {Lutz}(2009)}]{Hilt2009a}%
  \BibitemOpen
  \bibfield  {author} {\bibinfo {author} {\bibfnamefont {S.}~\bibnamefont
  {Hilt}}\ and\ \bibinfo {author} {\bibfnamefont {E.}~\bibnamefont {Lutz}},\
  }\href@noop {} {\bibfield  {journal} {\bibinfo  {journal} {Phys. Rev. A}\
  }\textbf {\bibinfo {volume} {79}},\ \bibinfo {pages} {010101} (\bibinfo
  {year} {2009})}\BibitemShut {NoStop}%
\bibitem [{\citenamefont {Campisi}\ \emph {et~al.}(2009)\citenamefont
  {Campisi}, \citenamefont {Talkner},\ and\ \citenamefont
  {H\"anggi}}]{Campisi2009}%
  \BibitemOpen
  \bibfield  {author} {\bibinfo {author} {\bibfnamefont {M.}~\bibnamefont
  {Campisi}}, \bibinfo {author} {\bibfnamefont {P.}~\bibnamefont {Talkner}}, \
  and\ \bibinfo {author} {\bibfnamefont {P.}~\bibnamefont {H\"anggi}},\
  }\href@noop {} {\bibfield  {journal} {\bibinfo  {journal} {Phys. Rev. Lett.}\
  }\textbf {\bibinfo {volume} {102}},\ \bibinfo {pages} {210401} (\bibinfo
  {year} {2009})}\BibitemShut {NoStop}%
\bibitem [{\citenamefont {Esposito}\ \emph
  {et~al.}(2015{\natexlab{b}})\citenamefont {Esposito}, \citenamefont {Ochoa},\
  and\ \citenamefont {Galperin}}]{Esposito2015}%
  \BibitemOpen
  \bibfield  {author} {\bibinfo {author} {\bibfnamefont {M.}~\bibnamefont
  {Esposito}}, \bibinfo {author} {\bibfnamefont {M.~A.}\ \bibnamefont {Ochoa}},
  \ and\ \bibinfo {author} {\bibfnamefont {M.}~\bibnamefont {Galperin}},\
  }\href@noop {} {\bibfield  {journal} {\bibinfo  {journal} {Phys. Rev. B}\
  }\textbf {\bibinfo {volume} {91}},\ \bibinfo {pages} {115417} (\bibinfo
  {year} {2015}{\natexlab{b}})}\BibitemShut {NoStop}%
\bibitem [{\citenamefont {Ludovico}\ \emph {et~al.}()\citenamefont {Ludovico},
  \citenamefont {Battista}, \citenamefont {von Oppen},\ and\ \citenamefont
  {Arrachea}}]{Ludovico2015}%
  \BibitemOpen
  \bibfield  {author} {\bibinfo {author} {\bibfnamefont {M.~F.}\ \bibnamefont
  {Ludovico}}, \bibinfo {author} {\bibfnamefont {F.}~\bibnamefont {Battista}},
  \bibinfo {author} {\bibfnamefont {F.}~\bibnamefont {von Oppen}}, \ and\
  \bibinfo {author} {\bibfnamefont {L.}~\bibnamefont {Arrachea}},\ }\href@noop
  {} {\ }\Eprint {http://arxiv.org/abs/1506.08617} {arXiv:1506.08617}
  \BibitemShut {NoStop}%
\bibitem [{\citenamefont {Esposito}\ \emph {et~al.}(2010)\citenamefont
  {Esposito}, \citenamefont {Lindenberg},\ and\ \citenamefont {Van~den
  Broeck}}]{Esposito2010a}%
  \BibitemOpen
  \bibfield  {author} {\bibinfo {author} {\bibfnamefont {M.}~\bibnamefont
  {Esposito}}, \bibinfo {author} {\bibfnamefont {K.}~\bibnamefont
  {Lindenberg}}, \ and\ \bibinfo {author} {\bibfnamefont {C.}~\bibnamefont
  {Van~den Broeck}},\ }\href@noop {} {\bibfield  {journal} {\bibinfo  {journal}
  {New J. Phys.}\ }\textbf {\bibinfo {volume} {12}},\ \bibinfo {pages} {013013}
  (\bibinfo {year} {2010})}\BibitemShut {NoStop}%
\bibitem [{\citenamefont {Deffner}\ and\ \citenamefont
  {Lutz}(2011)}]{Deffner2011a}%
  \BibitemOpen
  \bibfield  {author} {\bibinfo {author} {\bibfnamefont {S.}~\bibnamefont
  {Deffner}}\ and\ \bibinfo {author} {\bibfnamefont {E.}~\bibnamefont {Lutz}},\
  }\href@noop {} {\bibfield  {journal} {\bibinfo  {journal} {Phys. Rev. Lett.}\
  }\textbf {\bibinfo {volume} {107}},\ \bibinfo {pages} {140404} (\bibinfo
  {year} {2011})}\BibitemShut {NoStop}%
\bibitem [{\citenamefont {Bode}\ \emph {et~al.}(2011)\citenamefont {Bode},
  \citenamefont {Viola~Kusminskiy}, \citenamefont {Egger},\ and\ \citenamefont
  {von Oppen}}]{BodePRB}%
  \BibitemOpen
  \bibfield  {author} {\bibinfo {author} {\bibfnamefont {N.}~\bibnamefont
  {Bode}}, \bibinfo {author} {\bibfnamefont {S.}~\bibnamefont
  {Viola~Kusminskiy}}, \bibinfo {author} {\bibfnamefont {R.}~\bibnamefont
  {Egger}}, \ and\ \bibinfo {author} {\bibfnamefont {F.}~\bibnamefont {von
  Oppen}},\ }\href@noop {} {\bibfield  {journal} {\bibinfo  {journal} {Phys.
  Rev. Lett.}\ }\textbf {\bibinfo {volume} {107}},\ \bibinfo {pages} {036804}
  (\bibinfo {year} {2011})}\BibitemShut {NoStop}%
\bibitem [{\citenamefont {Bode}\ \emph {et~al.}(2012)\citenamefont {Bode},
  \citenamefont {Viola~Kusminskiy}, \citenamefont {Egger},\ and\ \citenamefont
  {von Oppen}}]{bode2012current}%
  \BibitemOpen
  \bibfield  {author} {\bibinfo {author} {\bibfnamefont {N.}~\bibnamefont
  {Bode}}, \bibinfo {author} {\bibfnamefont {S.}~\bibnamefont
  {Viola~Kusminskiy}}, \bibinfo {author} {\bibfnamefont {R.}~\bibnamefont
  {Egger}}, \ and\ \bibinfo {author} {\bibfnamefont {F.}~\bibnamefont {von
  Oppen}},\ }\href@noop {} {\bibfield  {journal} {\bibinfo  {journal}
  {Beilstein J. Nanotechnol.}\ }\textbf {\bibinfo {volume} {3}},\ \bibinfo
  {pages} {144} (\bibinfo {year} {2012})}\BibitemShut {NoStop}%
\bibitem [{\citenamefont {Esposito}\ \emph {et~al.}()\citenamefont {Esposito},
  \citenamefont {Ochoa},\ and\ \citenamefont {Galperin}}]{Esposito2014}%
  \BibitemOpen
  \bibfield  {author} {\bibinfo {author} {\bibfnamefont {M.}~\bibnamefont
  {Esposito}}, \bibinfo {author} {\bibfnamefont {M.~A.}\ \bibnamefont {Ochoa}},
  \ and\ \bibinfo {author} {\bibfnamefont {M.}~\bibnamefont {Galperin}},\
  }\href@noop {} {\ }\Eprint {http://arxiv.org/abs/1408.3608} {arXiv:1408.3608}
  \BibitemShut {NoStop}%
\bibitem [{Note2()}]{Note2}%
  \BibitemOpen
  \bibinfo {note} {Note that, as we work in the grand canonical ensemble
  framework, the metal lead in our 'world' is assumed to be weakly open to an
  equilibrium bath of given temperature and electronic chemical
  potential.}\BibitemShut {Stop}%
\bibitem [{\citenamefont {Ajisaka}\ \emph {et~al.}(2012)\citenamefont
  {Ajisaka}, \citenamefont {Barra}, \citenamefont {Mej{\'{\i}}a-Monasterio},\
  and\ \citenamefont {Prosen}}]{Ajisaka2012}%
  \BibitemOpen
  \bibfield  {author} {\bibinfo {author} {\bibfnamefont {S.}~\bibnamefont
  {Ajisaka}}, \bibinfo {author} {\bibfnamefont {F.}~\bibnamefont {Barra}},
  \bibinfo {author} {\bibfnamefont {C.}~\bibnamefont
  {Mej{\'{\i}}a-Monasterio}}, \ and\ \bibinfo {author} {\bibfnamefont
  {T.}~\bibnamefont {Prosen}},\ }\href@noop {} {\bibfield  {journal} {\bibinfo
  {journal} {Phys. Rev. B}\ }\textbf {\bibinfo {volume} {86}},\ \bibinfo
  {pages} {125111} (\bibinfo {year} {2012})}\BibitemShut {NoStop}%
\bibitem [{Note3()}]{Note3}%
  \BibitemOpen
  \bibinfo {note} {The velocity of the level is measured by $\protect
  \mathaccentV {dot}05F{\varepsilon }_d / \Gamma $ and the detailed condition
  for the process being quasistatic depends on whether $k_B T< \Gamma $ or $k_B
  T> \Gamma $. In these limits, one obtains the conditions $\protect
  \mathaccentV {dot}05F{\varepsilon }_{d} / \Gamma \ll \Gamma $ and $\protect
  \mathaccentV {dot}05F{\varepsilon }_{d} / \Gamma \ll k_B T $,
  respectively.}\BibitemShut {Stop}%
\bibitem [{\citenamefont {Tsvelick}\ and\ \citenamefont
  {Wiegmann}(1983)}]{Tsvelick1983}%
  \BibitemOpen
  \bibfield  {author} {\bibinfo {author} {\bibfnamefont {A.}~\bibnamefont
  {Tsvelick}}\ and\ \bibinfo {author} {\bibfnamefont {P.}~\bibnamefont
  {Wiegmann}},\ }\href@noop {} {\bibfield  {journal} {\bibinfo  {journal} {Adv.
  Phys.}\ }\textbf {\bibinfo {volume} {32}},\ \bibinfo {pages} {453} (\bibinfo
  {year} {1983})}\BibitemShut {NoStop}%
\bibitem [{\citenamefont {Nghiem}\ \emph {et~al.}()\citenamefont {Nghiem},
  \citenamefont {Kennes}, \citenamefont {Kl{\"{o}}ckner}, \citenamefont
  {Meden},\ and\ \citenamefont {Costi}}]{Nghiem2016}%
  \BibitemOpen
  \bibfield  {author} {\bibinfo {author} {\bibfnamefont {H.~T.~M.}\
  \bibnamefont {Nghiem}}, \bibinfo {author} {\bibfnamefont {D.~M.}\
  \bibnamefont {Kennes}}, \bibinfo {author} {\bibfnamefont {C.}~\bibnamefont
  {Kl{\"{o}}ckner}}, \bibinfo {author} {\bibfnamefont {V.}~\bibnamefont
  {Meden}}, \ and\ \bibinfo {author} {\bibfnamefont {T.~A.}\ \bibnamefont
  {Costi}},\ }\href@noop {} {\ }\Eprint {http://arxiv.org/abs/1601.04080}
  {arXiv:1601.04080} \BibitemShut {NoStop}%
\bibitem [{\citenamefont {Weiss}(2008)}]{Weiss1999}%
  \BibitemOpen
  \bibfield  {author} {\bibinfo {author} {\bibfnamefont {U.}~\bibnamefont
  {Weiss}},\ }\href@noop {} {\bibfield  {journal} {\bibinfo  {journal}
  {\textit{Quantum Dissipative Systems}, World Scientific, 3rd Edition}\ }
  (\bibinfo {year} {2008})}\BibitemShut {NoStop}%
\bibitem [{\citenamefont {Kita}(2010)}]{Kita2010}%
  \BibitemOpen
  \bibfield  {author} {\bibinfo {author} {\bibfnamefont {T.}~\bibnamefont
  {Kita}},\ }\href@noop {} {\bibfield  {journal} {\bibinfo  {journal} {Prog.
  Theor. Phys.}\ }\textbf {\bibinfo {volume} {123}},\ \bibinfo {pages} {581}
  (\bibinfo {year} {2010})}\BibitemShut {NoStop}%
\bibitem [{\citenamefont {Jauho}\ \emph {et~al.}(1994)\citenamefont {Jauho},
  \citenamefont {Wingreen},\ and\ \citenamefont
  {Meir}}]{JauhoTime-DependentTransp}%
  \BibitemOpen
  \bibfield  {author} {\bibinfo {author} {\bibfnamefont {A.-P.}\ \bibnamefont
  {Jauho}}, \bibinfo {author} {\bibfnamefont {N.~S.}\ \bibnamefont {Wingreen}},
  \ and\ \bibinfo {author} {\bibfnamefont {Y.}~\bibnamefont {Meir}},\
  }\href@noop {} {\bibfield  {journal} {\bibinfo  {journal} {Phys. Rev. B}\
  }\textbf {\bibinfo {volume} {50}},\ \bibinfo {pages} {5528} (\bibinfo {year}
  {1994})}\BibitemShut {NoStop}%
\bibitem [{\citenamefont {Stefanucci}\ and\ \citenamefont
  {Almbladh}(2004)}]{Stefanucci2004}%
  \BibitemOpen
  \bibfield  {author} {\bibinfo {author} {\bibfnamefont {G.}~\bibnamefont
  {Stefanucci}}\ and\ \bibinfo {author} {\bibfnamefont {C.-O.}\ \bibnamefont
  {Almbladh}},\ }\href@noop {} {\bibfield  {journal} {\bibinfo  {journal}
  {Phys. Rev. B}\ }\textbf {\bibinfo {volume} {69}},\ \bibinfo {pages} {195318}
  (\bibinfo {year} {2004})}\BibitemShut {NoStop}%
\bibitem [{\citenamefont {Haug}\ and\ \citenamefont
  {Jauho}(1996)}]{haug1996quantum}%
  \BibitemOpen
  \bibfield  {author} {\bibinfo {author} {\bibfnamefont {H.}~\bibnamefont
  {Haug}}\ and\ \bibinfo {author} {\bibfnamefont {A.}~\bibnamefont {Jauho}},\
  }\href@noop {} {\bibfield  {journal} {\bibinfo  {journal} {\textit{Quantum
  Kinetics in Transport and Optics of Semiconductors}, Springer}\ } (\bibinfo
  {year} {1996})}\BibitemShut {NoStop}%
\bibitem [{Note4()}]{Note4}%
  \BibitemOpen
  \bibinfo {note} {Eq.\ \protect \textup {\hbox {\mathsurround \z@ \protect
  \normalfont (\ignorespaces \ref {QBoltzmannEq}\unskip \@@italiccorr )}}
  differs from Eq.\ (4.19) in Ref.\ \protect \rev@citealpnum {Kita2010}, used
  in Ref.\ \protect \rev@citealpnum {EspositoQThermo}, in the second Poisson
  bracket on the left, as our expression involves $f$ rather than $\phi $. We
  believe that our form is correct, but in any case both forms are equivalent
  up to the first order in velocity considered here and both lead to the same
  solution for $\phi $ Eq.\ \protect \textup {\hbox {\mathsurround \z@ \protect
  \normalfont (\ignorespaces \ref {PhiQBoltzmann}\unskip \@@italiccorr
  )}}.}\BibitemShut {Stop}%
\end{thebibliography}
\end{document}